\documentclass[twocolumn,superscriptaddress,showpacs,nofootinbib,aps,prd,preprintnumbers,floatfix]{revtex4-1}
\usepackage{lmodern}
\usepackage[T1]{fontenc}
\usepackage[utf8]{inputenc}
\usepackage{amstext,amssymb,amsmath}
\usepackage[table]{xcolor}
\usepackage{multirow}
\usepackage{graphicx}
\usepackage{slashed}
\usepackage{enumitem}
\usepackage[normalem]{ulem}
\usepackage{comment}
\usepackage{cases}

\definecolor{LightGray}{gray}{0.91}
\definecolor{LightBlue}{rgb}{0.87, 0.94, 1}

\usepackage[pdfusetitle,unicode]{hyperref}
\hypersetup{       
	colorlinks=false,
	breaklinks=true,
	linkcolor=green,
	linkbordercolor=green,
	anchorcolor=black,
	filecolor=cyan,
	filebordercolor=cyan,
	menucolor=blue, 
	menubordercolor=blue,
	runcolor=cyan,
	runbordercolor=cyan,
	citecolor=blue,
	citebordercolor=blue,
	urlcolor=white,
	urlbordercolor=white,
	pdfborderstyle={/S/U/W 0.5}, 
	linktocpage=true, 
}

\usepackage[capitalise]{cleveref} 
\crefname{equation}{Eq.}{Eqs.}
\Crefname{equation}{Equation}{Equations} 

\crefname{table}{Tab.}{Tabs.}
\Crefname{table}{Table}{Tables}

\crefname{figure}{Fig.}{Figs.}
\Crefname{figure}{Figure}{Figures}

\crefname{chapter}{Chap.}{Chaps.}
\Crefname{chapter}{Chapter}{Chapters}

\crefname{section}{Sec.}{Secs.}
\Crefname{section}{Section}{Sections}

\crefname{appendix}{App.}{Apps.}
\Crefname{appendix}{Appendix}{Appendices}

\usepackage{multirow}
\usepackage{epsfig,verbatim,mathrsfs,array,layout,textcomp,latexsym}

\usepackage{tikz}
\usepackage[compat=1.1.0]{tikz-feynman}
\tikzfeynmanset{warn luatex=false}
\usetikzlibrary{tikzmark,fit,arrows,matrix,positioning,fit,calc}

\allowdisplaybreaks[1]

\renewcommand{\(}{\left(}
\renewcommand{\)}{\right)}
\renewcommand{\{}{\left\lbrace}
\renewcommand{\}}{\right\rbrace}

\newcommand{\order}[1]{\mathcal{O}\({#1}\)}

\newcommand{\phm}{\phantom{-}} 

\newcommand{\alphae}{\alpha_\text{e}}
\newcommand{\gfermi}{G_\text{F}}

\newcommand{\gev}{\text{GeV}}

\newcommand{\wilsonC}[1]{c_{#1}}
\newcommand{\wilsonCp}[1]{c_{#1}^\prime}

\usepackage{xspace}
\def\python     {\mbox{\textsc{Python}}\xspace}
\def\flavio		{\mbox{\texttt{flavio}}\xspace}
\def\iminuit	{\mbox{\texttt{iminuit}}\xspace}
\def\ALmigrad   {\mbox{\texttt{MIGRAD}}\xspace}
\def\ALminos   {\mbox{\texttt{MINOS}}\xspace}

\makeatletter
    \def\CT@@do@color{%
      \global\let\CT@do@color\relax
            \@tempdima\wd\z@
            \advance\@tempdima\@tempdimb
            \advance\@tempdima\@tempdimc
    \advance\@tempdimb\tabcolsep
    \advance\@tempdimc\tabcolsep
    \advance\@tempdima2\tabcolsep
            \kern-\@tempdimb
            \leaders\vrule
                    \hskip\@tempdima\@plus  1fill
            \kern-\@tempdimc
            \hskip-\wd\z@ \@plus -1fill }
    \makeatother

\begin{document}
\preprint{DO-TH 21/30}

\title{Model-independent analysis of $\boldsymbol{b  \to d}$ processes}

\author{Rigo Bause}
\email{rigo.bause@tu-dortmund.de}
\author{Hector Gisbert}
\email{hector.gisbert@tu-dortmund.de}
\author{Marcel Golz}
\email{marcel.golz@tu-dortmund.de}
\author{Gudrun Hiller}
\email{ghiller@physik.uni-dortmund.de}
\affiliation{TU Dortmund University, Department of Physics, Otto-Hahn-Str.4, D-44221 Dortmund, Germany}

\begin{abstract}
We perform a model-independent analysis of $|\Delta b|=|\Delta d|=1$ processes to test the standard model and probe flavor patterns of new physics. 
Constraints on Wilson coefficients are obtained from global fits to $B^+ \to \pi^+ \,\mu^+\mu^-$, $B^0_s\to \bar{K}^{*0}\, \mu^+\mu^-$, $B^0\to\mu^+\mu^-$, and radiative $B \to X_d \,\gamma$ decays data. 
The fits are consistent with the standard model but leave sizable room for new physics. 
Besides higher-statistics measurements and  more  data in theory-friendly bins of the dilepton mass, further complementary observables such as angular distributions of $B_s^0 \to  \bar{K}^{*0}\,\ell^+ \ell^-$, $B \to \rho\,\ell^+ \ell^-$  or the baryonic modes $\Xi_b \to \Sigma\,\ell^+ \ell^-$, $\Omega_b^- \to \Xi^- \, \ell^+ \ell^-$ are necessary to resolve the significant degeneracy in the fit for the semileptonic four-fermion operators. 
Assuming minimal quark flavor violation,  the $b \to s$ global fit implies  tight constraints on the  $b \to d$ couplings, and hence allows to test this paradigm with improved data. 
Another benefit from $|\Delta b|=|\Delta d|=1$ processes is to shed light on the  $B$-anomalies in $|\Delta b|=|\Delta s|=1$ modes from a new angle.
Specifically, studies of lepton flavor-specific and dineutrino modes are informative on the lepton flavor structure. Rare $b \to d \, \ell \ell, \nu \bar \nu$ decays can be studied at high luminosity flavor facilities LHCb, Belle II, and a future $Z$-factory.
\end{abstract}

\maketitle
\tableofcontents
\flushbottom

\section{Introduction}

Flavor-changing neutral current (FCNC) transitions arise
in the standard model (SM) at the quantum level, and are sensitive to 
new physics (NP) and its flavor structure. 
Rare radiative  decays $b\to q\,\gamma$ and semileptonic ones $b\to q\,\ell^-\ell^+$ with $q=s,d$ and $\ell=e,\mu,\tau$ are such promising probes,  
allowing also to test approximate symmetries of the 
SM such as lepton flavor universality.

Especially the  $b\to s\,\mu^+\mu^-$ transitions have been analyzed over the past two decades with vigor, revealing an intriguing picture of 
discrepancies with SM predictions, 
referred to as the flavor anomalies, e.g.~\cite{Bifani:2018zmi,Albrecht:2021tul,London:2021lfn}:  
{\it i)} Branching ratios are below the SM values.
{\it ii)} Angular distributions provide theoretically cleaner observables and give more than $4 \sigma$ deviation from the SM in global fits, 
recently~\cite{Alguero:2021anc,Kriewald:2021hfc,Geng:2021nhg,Bause:2021cna}.
Electron-muon universality violation has been evidenced by LHCb in  $R_K$, the ratio 
of $B\to K \mu \mu$ to $B \to K ee$ branching fractions \cite{LHCb:2021trn}, strengthening the trend in measurements of $R_K$-like ratios \cite{Hiller:2003js} with
$b$-decays into  other strange hadrons. Interestingly, a very recent experimental update  of $R_K$ and $R_{K^*}$~\cite{LHCb:2022zom,LHCb:2022qnv} revealed  consistency with the SM.
Since data {\it i)} and  {\it ii)} can be  explained  by NP in semileptonic  $|\Delta b|=|\Delta s|=1$ four-fermion operators with coupling to muons,
the nil return of electron-muon non-universality in the new LHCb  results \cite{LHCb:2022zom,LHCb:2022qnv}   suggests discrepancies with the SM in $b \to s e^+ e^-$ modes, specifically similarly  reduced  branching ratios and distorted angular distributions.
While this  points to NP  
in flavorful processes,
further scrutiny is required before firm conclusions can be drawn. 
This includes also cross-checks with other sectors.
 
In this work, 
we perform a model-independent analysis of $|\Delta b|=|\Delta d|=1$ processes.
Such modes are subject to a Cabibbo–Kobayashi–Maskawa (CKM) suppression 
relative to $|\Delta b|=|\Delta s|=1$ ones,
indicating branching ratios 
smaller by two powers of the Wolfenstein parameter,
$\mathcal{B}(b\to d)\sim 0.04\cdot \mathcal{B}(b\to s)$.
Only branching ratios of rare $b \to d\,\mu^+ \mu^-, \gamma$ decays have been measured: $B^+\to\pi^+\,\mu^+\mu^-$ ~\cite{LHCb:2015hsa},
$B^0\to\mu^+\mu^-$~\cite{LHCb:2021awg}, 
and the first evidence of $B_s^0\to \bar{K}^{*0} \mu^+\mu^-$~\cite{LHCb:2018rym},
as well as $B\to X_d\gamma$~\cite{Misiak:2015xwa,BaBar:2010vgu}.
Our goal is to take this  data set 
and extract information on Wilson coefficients in the weak effective theory (EFT) from global fits,
for earlier analyses see Refs.~\cite{Du:2015tda,Ali:2013zfa,Rusov:2019ixr}.

In addition to benefitting from correlations among several $b \to d$ observables, 
model-independent EFT-interpretations 
also allow for probing the quark and lepton flavor structure of NP.
Analyses within the SM effective theory (SMEFT) \cite{Buchmuller:1985jz,Grzadkowski:2010es}  have provided directions to shed light on the flavor anomalies and
beyond by combining top, beauty, charm, and kaon,   
and charged dilepton and dineutrino data~\cite{Bissmann:2020mfi,Bause:2020auq,Bruggisser:2021duo}.
The results of this work are providing input for the connection between third- and first-generation quark FCNCs,
their link to the third- and second-generation quark FCNCs, 
and tests of lepton universality in $ b \to d\,\nu \bar \nu$ decays. Specifically, we give details and updates to the analysis of \cite{Bause:2021cna}.

The paper is organized as follows: 
In \cref{sec:theory}, 
we give the effective theory framework and the $|\Delta b|=|\Delta d|=1$ rare decay observables used in the analysis. 
Set-up and results of the global fits are presented in \cref{sec:global}. 
We conclude in \cref{sec:con}.
Covariance matrices are given in 
\cref{app:covarianceObs}.
Uncertainty estimates of the tails of charmonia on the $B_s^0\to \bar{K}^{*0} \mu^+\mu^-$ branching ratios using LHCb's cuts~\cite{LHCb:2018rym} are given in 
\cref{app:resonances}.

\section{Effective theory framework}\label{sec:theory}

In \cref{sec:mia}, we introduce the weak effective theory framework to study $|\Delta b|=|\Delta d|=1$ transitions. The SM predictions for (semi-)leptonic and radiative observables, as well as its semi-analytical expressions, where NP effects are included model-independently, are presented in \cref{sec:obs}.

\subsection{Weak effective Hamiltonian}\label{sec:mia}

Rare $b\to d\,\mu^+\mu^-$ transitions can be described by the following effective Hamiltonian
\begin{align}\label{eq:hamiltonianNP}
    \mathcal{H}_{\text{eff}}&=-\frac{4\,\gfermi}{\sqrt{2}}\left(\lambda_t^{(d)}\,\mathcal{H}_{\text{eff}}^{(t)}\,+\,\lambda_u^{(d)}\,\mathcal{H}_{\text{eff}}^{(u)}\right)\,+\,\text{h.c.}~,
\end{align}
with
\begin{align}
    \mathcal{H}_{\text{eff}}^{(t)}&\,=\,\wilsonC{1}\,\mathcal{O}_1^{c}\,+\,\wilsonC{2}\,\mathcal{O}_2^{c}\nonumber\\
    &\,+\,\sum_{i=3}^6 \wilsonC{i}\,\mathcal{O}_i\,+ \sum_{i=7}^{10}\left(\wilsonC{i}\,\mathcal{O}_i+\wilsonCp{i}\,\mathcal{O}_i^\prime\right)~,\\
    \mathcal{H}_{\text{eff}}^{(u)}&\,=\,\wilsonC{1}\,(\mathcal{O}_1^{c}-\mathcal{O}_1^{u})\,+\,\wilsonC{2}\,(\mathcal{O}_2^{c}-\mathcal{O}_2^{u})~,
\end{align}
and  the CKM factors $\lambda_i^{(d)}=V_{id}^* V_{ib}$. We consider NP effects in the following dimension-six operators
\begin{align}
    \mathcal{O}_7^{(\prime)} &=\frac{e}{16\pi^2}m_b\left(\bar{d}_{L(R)}\sigma^{\alpha\beta}b_{R(L)}\right)F_{\alpha\beta}~,\label{eq:O7}\\
    \mathcal{O}_8^{(\prime)} &=\frac{g_s}{16\pi^2}m_b\left(\bar{d}_{L(R)} \sigma^{\alpha\beta} T^a b_{R(L)}\right)G^{a}_{\alpha\beta}~,\label{eq:O8}\\
    \mathcal{O}_9^{(\prime)} &=\frac{\alphae}{4\pi} \left(\bar{d}_{L(R)}\gamma_\alpha b_{L(R)}\right)\,\left(\bar{\mu}\gamma^\alpha\mu\right)~,\label{eq:O9}\\
    \mathcal{O}_{10}^{(\prime)} &=\frac{\alphae}{4\pi}  \left(\bar{d}_{L(R)}\gamma_\alpha b_{L(R)}\right)\,\left(\bar{\mu}\gamma^\alpha\gamma^5\mu\right)~.\label{eq:O10}
\end{align} 
Here, $\alphae=e^2/(4 \pi)$ ($\gfermi$) denotes the fine structure (Fermi's) constant,  $g_s$ is the strong QCD coupling, and  $F_{\alpha\beta}$, $G^{a}_{\alpha\beta}$ are the electromagnetic, chromomagnetic field strength tensors, respectively. The $T^a$ with $a=1,...,8$ are the generators of the $SU(3)_C$ group, and $\sigma^{\alpha\beta}=\frac{i}{2}[\gamma^\alpha,\gamma^\beta]$, and $L=(1-\gamma_5)/2, R=(1+\gamma_5)/2$ are projectors on left-, right-handed chirality. The Wilson coefficients 
\begin{align} \nonumber
c_i=C_i^{\text{SM}}+C_i\, , \quad c_i^\prime =C_i^{\prime}~,
\end{align}
encode the dynamics of the heavy degrees of freedom from  the SM (the top quark, the $W^\pm$ and $Z^0$-bosons, and the Higgs) and NP in  $C_i^{\text{SM}(\prime)}$ and $C_i^{(\prime)}$, respectively. Here, primed operators $\mathcal{O}_{7,8,9,10}^{\prime}$ are the  helicity-flipped counter-parts of $\mathcal{O}_{7,8,9,10}$. At the bottom-mass scale $\mu_b\approx m_b$ holds in the SM at next-to-next-to-leading order (NNLO) accuracy, $C_7^{\text{SM}}(\mu_b)=-0.30$, $C_8^{\text{SM}}(\mu_b)=-0.15$, $C_9^{\text{SM}}(\mu_b)=4.12$ and $C_{10}^{\text{SM}}(\mu_b)=-4.18$~\cite{Ali:2013zfa}. 
In the SM, the primed coefficients receive a suppression by light down-quark to $b$-quark mass relative to the unprimed ones, $C_i^{\text{SM} \prime} =(m_d/m_b)\, C_i^{\text{SM}}$ and can be safely neglected, in addition to (pseudo-)scalar and (pseudo-)tensor operators. In this work, we do not include the latter operators, but we work out limits on (pseudo-)scalar ones from the branching ratio of $B^0\to\mu^+\mu^-$ decay in \cref{sec:Btomumu}.

We assume that the four-quark operators  $\mathcal{O}_{1,2}^{q} \sim (\bar d \gamma_\mu b)( \bar q \gamma^\mu q), q=u,c$, and QCD penguins $O_{3 ...6}$  receive SM contributions only. The matrix elements of the four-quark operators are absorbed into ``effective''  coefficients $C_{7,9,10}^{\text{eff}}(q^2)$ of the operators $\mathcal{O}_{7,9,10}$\,, 
see Refs.~\cite{Bobeth:1999mk,Asatrian:2001de,Asatryan:2001zw,Ali:2002jg,Asatrian:2003vq,deBoer:2017way},
\begin{align}  \nonumber
 C_{i}^{\text{eff}}(q^2)=\wilsonC{i}+ \text{4-quark contributions} \, , 
\end{align}
which depend in general on the dilepton invariant mass squared denoted by $q^2$.

\subsection{Rare $b \to d$ decay observables} \label{sec:obs}

We introduce the $|\Delta b|=|\Delta d|=1$ observables taken into account in the global fits presented in \cref{sec:global}. Formulas are given in or are adapted from Refs.~\cite{Ali:2013zfa,Ali:1999mm,Hurth:2003dk}. 

\subsubsection{\texorpdfstring{$B^+\to \pi^+\mu^+\mu^-$}{B+ to pi+ mumu}}

The differential branching ratio of $B^+\to \pi^+ \mu^+\mu^-$ decays can be written as~\cite{Bobeth:2007dw,Ali:2013zfa}
\begin{align}
    \frac{\text{d}\,\mathcal{B}(B^+\to\pi^+\mu^+\mu^-)}{\text{d}q^2}=\mathcal{N}_{B^+}\,\lambda_{B\pi} ^{\frac{3}{2}}\,\mathcal{F}_{B\pi}(q^2)\,,
\end{align}
where $\mathcal{N}_{B}=2\,|V_{tb}V_{td}^*|^2\,\gfermi^2\,\alphae^2\,\tau_B/(3\,(4\,\pi)^5\,m_B^3)$, and $\lambda_{B\pi}(q^2)\equiv\lambda(m_B^2,m_\pi^2,q^2)$ 
is the usual Källén function with $\lambda(a,b,c)=a^2+b^2+c^2-2\,(ab+ac+bc)$. 
The dynamical function $\mathcal{F}_{B\pi}(q^2)$ 
\begin{align} \label{eq:functionBpi} \begin{split}
\mathcal{F}_{B\pi}(q^2)  = &\left | {C}_{9}^{\rm eff}(q^2) f_+(q^2) + 
\frac{2\, m_b\, {C}_{7}^{\rm eff}(q^2)  
\,f_T(q^2)}{m_B + m_\pi} \right |^2\\
+ &\left | {C}_{10}^{\rm eff}(q^2)  f_+(q^2)  \right |^2+\order{m_\mu^2}~,
\end{split}
\end{align}
encodes the dependence on the Wilson coefficients and the form factors (FFs). Here, in the low $q^2$ region, that is, in the kinematic region where the pion is energetic in the  $B$-meson center of mass frame we employ QCD factorization  
\cite{Beneke:2004dp}. We do not consider  very low $q^2$-bins near light resonances and those near the charmonium peaks.
At high $q^2$, i.e.,~low hadronic recoil we employ the OPE-framework of \cite{Grinstein:2004vb,Bobeth:2011nj}. As duality is expected to work better for larger intervals,
we use the largest available bin from that region.
In view of the presently sizable experimental uncertainites and the selected bins we refrain from studying the impact of non-FF contributions.
These include weak annihilation contributions, that are of importance at very low $q^2$~\cite{Hou:2014dza,Ali:2020tjy}.
In
\cref{eq:functionBpi} terms induced by a  finite muon mass are omitted for clarity, but are included in the numerical analysis.
The functions $f_+(q^2)$, $f_T(q^2)$ (and $f_0(q^2)$ when corrections $\order{m_\mu^2}$ are taken into account) denote the $B\to\pi$ transition FFs, and their determination requires the use of non-perturbative techniques. 
We employ the results of Ref.~\cite{Leljak:2021vte}, where a fit combining lattice data at high-$q^2$, and light-cone sum rules (LCSRs) data at low--$q^2$ has been performed. 
We also include the contributions from Ref.~\cite{deBoer:2017way} in the effective coefficients.

\begin{table}[ht!]
 \centering
 \begingroup
 \resizebox{0.47\textwidth}{!}{
 \renewcommand{\arraystretch}{1.55}
 \begin{tabular}{cccc}
  \hline
  \hline
  \rowcolor{LightBlue} 
  && &\\
  \rowcolor{LightBlue} 
  $k$&$[q^2_ {\text{min}},q^2_ {\text{max}}]$ &\multicolumn{2}{c}{$\mathcal{B}^{(B\pi)}_k$} \\
  \rowcolor{LightBlue} 
  & & SM & experiment \\
  \rowcolor{LightBlue} 
 & $[\gev^2]$ &$[10^{-9}\,\gev^{-2}]$ &$[10^{-9}\,\gev^{-2}]$\\
  \hline
  \rowcolor{LightGray} 
  1 &$[2,4]$ & $0.80 \pm 0.12\pm0.05\pm0.04$  & $0.62^{+0.39}_{-0.33} \pm 0.02$ \\
  \rowcolor{LightGray} 
  2 &$[4,6]$ & $0.81 \pm 0.12\pm0.05\pm0.05$ & $0.85^{+0.32}_{-0.27} \pm 0.02$ \\
  3&$[6,8]$ & $0.82 \pm 0.11\pm 0.05\pm 0.07$ & $0.66^{+0.30}_{-0.25} \pm 0.02$ \\
  4&$[11,12.5]$ & $0.82 \pm 0.09\pm 0.05\pm 0.09$ & $0.88^{+0.34}_{-0.29} \pm 0.03$ \\
  5&$[15,17]$ & $0.73 \pm 0.06\pm 0.04\pm 0.06$ & $0.63^{+0.24}_{-0.19} \pm 0.02$ \\
  6&$[17,19]$ & $0.67 \pm 0.05\pm 0.04\pm 0.05$ & $0.41^{+0.21}_{-0.17} \pm 0.01$ \\
  7&$[19,22]$ & $0.57 \pm 0.03\pm 0.03\pm 0.04$ & $0.38^{+0.18}_{-0.15} \pm 0.01$ \\
  8&$[22,25]$ & $0.35 \pm 0.02\pm 0.02\pm 0.02$ & $0.14^{+0.13}_{-0.09} \pm 0.01$ \\
  \rowcolor{LightGray} 
  9 &$[15,22]$ & $0.64 \pm 0.04\pm 0.04\pm 0.05$ & $0.47^{+0.12}_{-0.10} \pm 0.01$ \\
  \hline
  10&$[4 m_\mu^2,(m_{B^+}-m_{\pi^+})^2]$  & $17.9 \pm 1.9\pm 1.1\pm 1.5^\dagger~ \gev^{2}$  & $18.3 \pm 2.4 \pm 0.5~ \gev^{2}$\\
  \hline
  \hline
  \end{tabular}
  }
  \endgroup
  \caption{SM non-resonant binned branching fractions $\mathcal{B}^{(B\pi)}_k$ in units of $10^{-9}\,\gev^2$, as defined in \cref{eq:binned}. The SM predictions are given in the third column with their three main sources of uncertainty including FFs, CKM matrix elements, and $\mu_b$, respectively. The experimental values from LHCb~\cite{LHCb:2015hsa} are displayed in the last column with their statistical and systematic uncertainties, respectively. Further details on the uncertainties treatment are provided in \cref{app:covarianceObs}. The shaded rows indicate the bins included in the fit, see main text for details. $^\dagger$To allow  for a  direct comparison with the value provided by LHCb the last row ($k=10$) corresponds to the integrated branching ratio over the full $q^2$-region, $((m_{B^+}-m_{\pi^+})^2-4\, m_\mu^2)\cdot \mathcal{B}^{(B\pi)}_{10}$, rather than a bin-average as in \cref{eq:binned}.}
  \label{tab:Bpi}
\end{table}

\begin{table}[ht!]
 \centering
 \begingroup
  \resizebox{0.49\textwidth}{!}{
  \renewcommand{\arraystretch}{1.55}
 \begin{tabular}{cccccccccccccc}
  \hline
  \hline
  \rowcolor{LightBlue} 
  $k$ & $[q^2_ {\text{min}},q^2_ {\text{max}}]$ & $a_{1}^{(B\pi)}$ & $a_{2}^{(B\pi)}$ &$a_{3}^{(B\pi)}$ &$a_{4}^{(B\pi)}$ &$a_{5}^{(B\pi)}$ &$a_{6}^{(B\pi)}$ &$a_{7}^{(B\pi)}$ &$a_{8}^{(B\pi)}$ &$a_{9}^{(B\pi)}$ &$a_{10}^{(B\pi)}$ &$a_{11}^{(B\pi)}$ &$a_{12}^{(B\pi)}$ \\
  \rowcolor{LightBlue} 
 &  $[\gev^2]$ & & & & & & & & & & & &\\
  \hline
  \rowcolor{LightGray} 
  1 & $[2,4]$     & 8.03 &3.37 &0.26&1.90&-2.17& 0.81&0.006&0.26&0.26&0.12&0.92&0.07 \\
    \rowcolor{LightGray} 
  2  & $[4,6]$     & 8.08 & 3.38& 0.23& 1.91&-2.18&0.81&0.005&0.26&0.26&0.11&0.92&0.06\\
  3 & $[6,8]$     & 8.17&3.41&0.21&1.94&-2.18&0.80&0.005&0.26&0.26&0.10&0.92&0.06 \\
  4 & $[11,12.5]$ &  8.16&3.40&0.19&1.94&-2.11&0.77&0.004&0.25&0.25&0.08&0.88&0.05 \\
  5 & $[15,17]$   & 7.34&3.06&0.16&1.74&-1.92&0.71&0.003&0.23&0.23&0.07&0.81&0.04\\
  6 & $[17,19]$   & 6.74&2.82&0.14&1.59&-1.78&0.66&0.003&0.21&0.21&0.06&0.75&0.04 \\
  7 & $[19,22]$   &5.66&2.36&0.12&1.34&-1.50&0.56&0.002&0.18&0.18& 0.05&0.63&0.03 \\
  8 & $[22,25]$   & 3.54&1.47&0.07&0.83&-0.94&0.35&0.001&0.11&0.11&0.03&0.39&0.02 \\
  \rowcolor{LightGray} 
  9  & $[15,22]$   & 6.45&2.69&0.14&1.53&-1.70&0.63&0.003&0.20&0.20&0.06&0.72&0.03 \\
  \hline
  10 & $[4 m_\mu^2,(m_{B}-m_{\pi})^2]$  & 178.66&74.61&4.64&42.35&-47.05&17.44&0.10&5.62&5.62&2.06&19.79&1.17 \\
  \hline
  \hline
  \end{tabular}
  }
  \endgroup
  \caption{Central values of $a^{(B\pi)}_{i}$ in units of $10^{-10}$ from \cref{eq:BpiNP}. Entries of the last row ($k=10$) are in $\gev^2$. The shaded rows indicate the bins included in the fit, see main text. Further details about uncertainties and correlations are provided in \cref{app:covarianceObs}.
  }
  \label{tab:Bpiparameters}
\end{table}

The SM predictions of the binned branching fractions
\begin{align}\label{eq:binned}
    \mathcal{B}^{(B\pi)}_k=\frac{\int_{q^2_\text{min}}^{q^2_\text{max}} \text{d}q^2 \frac{\text{d}\,\mathcal{B}(B^+\to\pi^+\mu^+\mu^-)}{\text{d}q^2}}{q^2_\text{max}-q^2_\text{min}},\,\,k=[q^2_\text{min},q^2_\text{max}]~,
\end{align}
with their respective uncertainties are given in the third column of \cref{tab:Bpi} \footnote{Note that we use the following shorthand notation, {\it i.e.} $\mathcal{B}^{(B\pi)}_1$ for the bin $[2,\,4]\,\gev^2$, and so on.}. The first source of error corresponds to FFs, the second one to CKM matrix elements, and the last one reflects the fluctuations under changes of the short-distance scale $\mu_b$ in the range between $m_b/2$ and $2\,m_b$ in the coefficients $C_i^{\text{eff}}$. Effects from $J/\psi$ and $\psi^\prime$ resonances are not included. Further details about the uncertainty treatment can be found in \cref{app:covarianceObs}.

\begin{figure}[h!]
    \centering
    \includegraphics[width=\columnwidth]{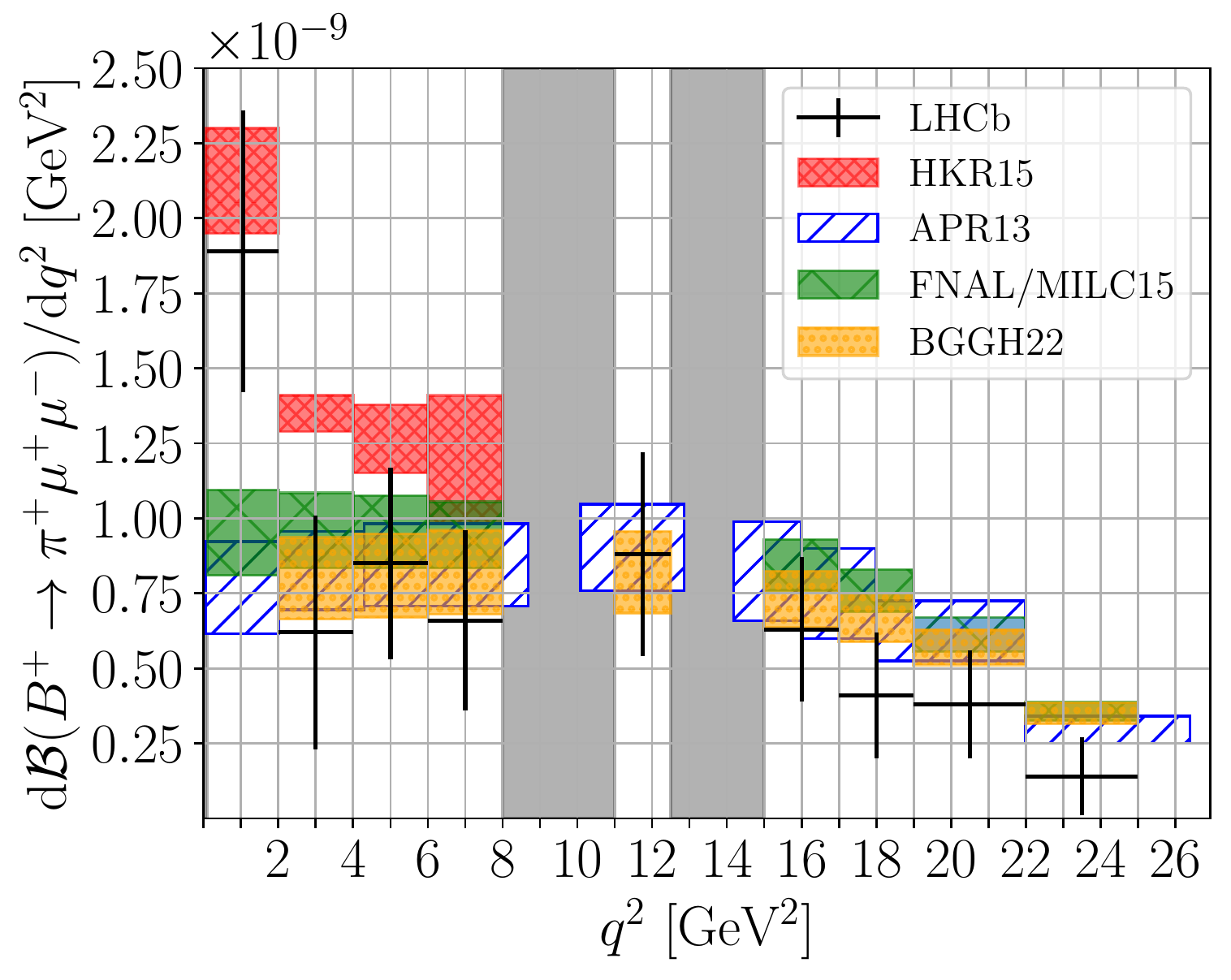}
    \caption{Measured  branching fraction of $B^+\to\pi^+\mu^+\mu^-$ in bins of dilepton invariant mass squared $q^2$ (black points) from LHCb~\cite{LHCb:2015hsa}.  SM predictions (this work, BGGH22, yellow) from \cref{tab:Bpi} are compared to those of  Refs.~\cite{Ali:2013zfa} (APR13, blue), \cite{Hambrock:2015wka} (HKR15, red) and~\cite{FermilabLattice:2015cdh} (FNAL/MILC15, green). Our prediction (yellow)  for the last bin is on top of the one from  FNAL/MILC15. Plot adapted
     from~\cite{LHCb:2015hsa}.  See text for details.}
\label{fig:lhcbplot}
\end{figure}

The measurements by LHCb~\cite{LHCb:2015hsa} (fourth column in \cref{tab:Bpi}) are in very good agreement with the SM predictions presented in the third column in \cref{tab:Bpi}. All bins are compatible within  $1\,\sigma$ except for the high-$q^2$ bin [22, 25]$\,\gev^2$, which agrees  at $1.6\,\sigma$. 
We do not provide theory predictions for the low-$q^2$ bin $[0.1,\, 2]\,\gev^2$, which suffers from hadronic uncertainties coming from the $\rho$, $\omega$ and $\phi$ resonances, $q^2\lesssim 1\,$GeV$^2$. In addition, regions around $q^2\approx9.5\,$\gev$^2$ and $q^2\approx 13.5\,$\gev$^2$ suffer from $J/\psi$ and $\psi^\prime$ resonances and their tails (gray bands in  \cref{fig:lhcbplot}), respectively. Therefore, in the global fits presented in \cref{sec:global} we \emph{only include} the theoretically clean bins  $k=1,2,9$, that are [2, 4] \gev$^2$,  [4, 6] \gev$^2$, and [15, 22] \gev$^2$, indicated by shaded rows in \cref{tab:Bpi,tab:Bpiparameters}. 

In \cref{fig:lhcbplot}, we compare the experimental branching fraction of $B^+\to\pi^+\mu^+\mu^-$ in bins of $q^2$ (black points) from LHCb~\cite{LHCb:2015hsa} with the SM predictions (BGGH22, this work, yellow) from \cref{tab:Bpi}. We compare our SM predictions to those available in the literature including Refs.~\cite{Ali:2013zfa} (APR13, blue), \cite{Hambrock:2015wka} (HKR15, red) and from lattice QCD calculations~\cite{FermilabLattice:2015cdh} (FNAL/MILC15, green). We observe very good agreement  except at low $q^2$ with the predictions of \cite{Hambrock:2015wka}, which are larger than the others. Ref.~\cite{Hambrock:2015wka}  includes non-FF contributions from non-local matrix elements from LCSR. On  top of that, the  branching ratios of \cite{Hambrock:2015wka} (red) are subject to an enhancement from the leading CKM-factor  by $8 \%$ and FFs by
almost $50 \%$ compared to our analysis (yellow). Taking this parametric effect into account addresses the bulk of the numerical differences between  the SM theory curves for 
$q^2 \gtrsim 2 \, \text{GeV}^2$.

In the presence of primed NP operators, \cref{eq:functionBpi} can be generalized by
\begin{align}\label{eq:subsBpi}
{C}_{i}^{\text{eff}}(q^2)\,\to\,{C}_{i}^{\text{eff}}(q^2)\,+\, C_i^\prime~, 
\end{align}
noting that due to spin parity of the external states and parity invariance of QCD the Wilson coefficients of unprimed and primed operators enter $B \to \pi \,\ell^+ \ell^-$ decays only by their exact sum, $C_{i^+}=C_i\, + \, C_i^\prime$. Taking into account \cref{eq:subsBpi}, the binned branching ratios \cref{eq:binned} can be compactly written as
\begin{align}\label{eq:BpiNP}
    \mathcal{B}_{k}^{(B\pi)}\,=\sum_{i=1}^{12}a_{i}^{(B\pi)}\,w_{i}^{(B\pi)}~,
\end{align}
with the following set of Wilson coefficients combinations
\begin{align} \label{eq:WCC}
    w^{(B\pi)}\,=\,&\left\lbrace 1,\, C_{7^+},\, C_{8^+},\, C_{9^+},\, C_{10^+},\, C_{7^+}^2, \,C_{8^+}^2,\, C_{9^+}^2,\right.\nonumber\\
    &\left.\, C_{10^+}^2,\, C_{7^+}\cdot C_{8^+},\, C_{7^+}\cdot C_{9^+},\, 
 C_{8^+}\cdot C_{9^+}\right\rbrace~.
\end{align}
The numerical values of $a^{(B\pi)}_{i}$ for different bins $k$ are given in \cref{tab:Bpiparameters}. Further information regarding uncertainties and correlations is provided in \cref{app:covarianceObs}.

\subsubsection{ \texorpdfstring{$B^0_s\to \bar{K}^{*0} \mu^+\mu^-$}{B0s to K*0mumu}}

The differential branching ratio of $B_s^0\to \bar{K}^{*0} \mu^+\mu^-$ decay can be expressed as~\cite{Ali:1999mm}
\begin{align}\label{eq:diffBtoKstar}
  &\frac{\text{d}\, \mathcal{B}(B_s^0\to \bar{K}^{*0} \mu^+\mu^-)}{\text{d} q^2}  =  \mathcal{N}_{B_s}\,\lambda_{B_s\bar{K}^*} ^{\frac{3}{2}}\,\mathcal{F}_{B_s\bar{K}^*}(q^2)\,,
\end{align}
where the dynamical function $\mathcal{F}_{B_s\bar{K}^*}(q^2)$ reads
\begin{align}\label{eq:functionBtoKstar}
    &\mathcal{F}_{B_s\bar{K}^*}(q^2)=\frac{1}{2}\left(|\mathscr{A}|^2+|\mathscr{E}|^2\right)\frac{q^2}{m_{B_s}^2}\nonumber\\
    &+\left(|\mathscr{B}|^2+|\mathscr{F}|^2\right)\left(\frac{1}{4}\frac{m_{B_s}^2}{m_{\bar{K}^*}^2}+3\frac{m_{B_s}^2 q^2}{\lambda_{B_s\bar{K}^*}}\right)\nonumber\\
    &+\frac{1}{2}\big(\text{Re}(\mathscr{B}\, \mathscr{C}^*)+\text{Re}(\mathscr{F}\, \mathscr{G}^*)\big)\bigg(\frac{q^2}{m_{\bar{K}^*}^2 }- \frac{m_{B_s}^2}{m_{\bar{K}^*}^2 }  + 1 \bigg)\nonumber\\
    &+\frac{1}{4}\left(|\mathscr{C}|^2+|\mathscr{G}|^2\right)\frac{\lambda_{B_s\bar{K}^*}}{m_B^2 m_{\bar{K}^*}^2}\,+\,\order{m_\mu^2}~.
\end{align}
Here, the functions $\mathscr{A},\,\mathscr{B},...$ contain the dependence on Wilson coefficients and FFs
\begin{align}\label{eq:BtoKstarABC}
    \mathscr{A}&=\frac{2}{1+\frac{m_{\bar{K}^*}}{m_{B_s}}}\, {C}_{9}^{\text{eff}}(q^2)\,V(q^2)+\frac{4 \,m_b}{q^2}\, {C}_{7}^\text{eff}(q^2)\,T_1(q^2)~,\nonumber\\
    \mathscr{B}&=\left(1+\frac{m_{\bar{K}^*}}{m_{B_s}}\right)\left[{C}_{9}^{\text{eff}}(q^2)\,A_1(q^2)\right.\nonumber\\
    &\left.+\frac{2 \,m_b}{q^2}\left(1-\frac{m_{\bar{K}^*}}{m_{B_s}}\right) {C}_{7}^\text{eff}(q^2)\,T_2(q^2)\right]~,\\
    \mathscr{C}&=\left(1-\frac{m_{\bar{K}^*}^2}{m_{B_s}^2}\right)^{-1}\left[\left(1-\frac{m_{\bar{K}^*}}{m_{B_s}}\right) {C}_{9}^{\text{eff}}(q^2)\,A_2(q^2)\right.\nonumber\\
    &\left.+2 \,m_b\, {C}_{7}^\text{eff}(q^2)\,\left(T_3(q^2)+\frac{m_{B_s}^2-m_{\bar{K}^*}^2}{q^2}T_2(q^2)\right)\right]~,\nonumber\\
    \mathscr{E}&=2\,\left(1+\frac{m_{\bar{K}^*}}{m_{B_s}}\right)^{-1}\,{C}_{10}^{\text{eff}}(q^2)\,V(q^2)~,\nonumber\\
    \mathscr{F}&=\,\left(1+\frac{m_{\bar{K}^*}}{m_{B_s}}\right)\,{C}_{10}^{\text{eff}}(q^2)\,A_1(q^2)~,\nonumber\\
    \mathscr{G}&=\left(1+\frac{m_{\bar{K}^*}}{m_{B_s}}\right)^{-1}\, {C}_{10}^{\text{eff}}(q^2)\,A_2(q^2)~,\nonumber
\end{align}
where $V(q^2)$, $A_{0,1,2}(q^2)$, $T_{1,2,3}(q^2)$ are the $B_s\to\bar{K}^*$ transition vector, axial and tensor FFs, respectively. We take them from Ref.~\cite{Gubernari:2018wyi}. As in \cref{eq:functionBpi}, we do not show the muon mass corrections. They are given  in Ref.~\cite{Ali:1999mm}, and are included in our analysis. Integrating \cref{eq:diffBtoKstar} over $[q^2_{\text{min}},q^2_{\text{max}}]=[0.1,19]\,\gev^2$, we find 
\begin{align}\label{eq:SMBtoKstar}
\begin{split}
\mathcal{B}(B_s^0\to \bar{K}^{*0} \mu^+\mu^-)_\text{SM}
=(46.0\pm 6.0)&\cdot 10^{-9}\\
=(46.0\pm 2.1 \pm 2.8 \pm 3.3 \pm 3.6)&\cdot 10^{-9}\,,
\end{split}
\end{align}
in agreement with the first experimental evidence from LHCb~\cite{LHCb:2018rym},
\begin{align}\label{eq:expBtoKstar}
\mathcal{B}(B_s^0\to \bar{K}^{*0} \mu^+\mu^-)_\text{exp}=(29\pm 11)\cdot 10^{-9}\,.
\end{align}
\Cref{eq:SMBtoKstar} highlights the four main sources of uncertainty:
FFs, CKM elements, the renormalization scale $\mu_b$, and the effect of the $J/\psi$ and $\psi^\prime$ resonances, respectively. The last uncertainty arises from the interpretation of  LHCb data~\cite{LHCb:2018rym} which encompasses the resonance regions. In \cref{app:resonances}, we provide details on the estimation of the last uncertainty, see \cref{eq:resonanceuncer}.

When chirality-flipped NP effects are turned on, thanks to Lorentz invariance and parity conservation of QCD, they enter either as a sum or difference with the unprimed coefficients, $C_{i^\pm}=C_i\,\pm\, C_i^\prime$. \Cref{eq:functionBtoKstar} can therefor be easily adapted through the following replacements in \cref{eq:BtoKstarABC}:
\begin{align}
\begin{split}
{C}_{i}^{\text{eff}}(q^2)\,V(q^2)&\to\,({C}_{i}^{\text{eff}}(q^2)\,+  C_i^\prime)\,V(q^2)~,\\ 
{C}_{i}^{\text{eff}}(q^2)\,A_{0,1,2}(q^2)&\to\,({C}_{i}^{\text{eff}}(q^2)\,- C_i^\prime)\,A_{0,1,2}(q^2)~,\\ 
{C}_{i}^{\text{eff}}(q^2)\,T_{1}(q^2)&\to\,({C}_{i}^{\text{eff}}(q^2)\,+ C_i^\prime)\,T_{1}(q^2)~,\\ 
{C}_{i}^{\text{eff}}(q^2)\,T_{2,3}(q^2)&\to\,({C}_{i}^{\text{eff}}(q^2)\,- C_i^\prime)\,T_{2,3}(q^2)~,
\end{split}
\end{align}
which result in
\begin{align}\label{eq:BsKstarPara}
    \mathcal{B}(B_s^0\to \bar{K}^{*0} \mu^+\mu^-)\,=\sum_{i=1}^{33}\,a^{(B_s\bar{K}^{*})}_{i}\,w^{(B_s\bar{K}^{*})}_{i}~,
\end{align}
with \footnote{\cref{eq:basisBsKstar} can be expressed as well in terms of $C_{i^\pm}$ through $C_{i}^{(\prime)}=\frac{1}{2}\left(C_{i^+}+(-)C_{i^-}\right)$~.}
\begin{align}\label{eq:basisBsKstar}
    w^{(B_s\bar{K}^{*})}_{i}\,=\,&\left\lbrace 1,\, C_{7},\, C_{8},\, C_{9},\, C_{10},\, C_{7}^2, \,C_{8}^2,\, C_{9}^2,\,C_{10}^2,\right.\nonumber\\
    &\left.\,C_{7}^\prime,\, C_{8}^\prime,\, C_{9}^\prime,\, C_{10}^\prime,\, (C_{7}^\prime)^2, \,(C_{8}^\prime)^2,\, (C_{9}^\prime)^2,\right.\nonumber\\
    &\left.\, (C_{10}^\prime)^2,\, C_{7}\cdot C_{7}^\prime,\,C_{8}\cdot C_{8}^\prime,\,C_{9}\cdot C_{9}^\prime,\,C_{10}\cdot C_{10}^\prime,\right.\nonumber\\
    &\left.\, C_{7}\cdot C_{8},\,C_{7}^\prime\cdot C_{8},\,C_{7}\cdot C_{8}^\prime,\,C_{7}^\prime\cdot C_{8}^\prime\right.\nonumber\\
    &\left.\, C_{7}\cdot C_{9},\,C_{7}^\prime\cdot C_{9},\,C_{7}\cdot C_{9}^\prime,\,C_{7}^\prime\cdot C_{9}^\prime\right.\nonumber\\
    &\left.\, C_{8}\cdot C_{9},\,C_{8}^\prime\cdot C_{9},\,C_{8}\cdot C_{9}^\prime,\,C_{8}^\prime\cdot C_{9}^\prime\right\rbrace \,.
\end{align}
The values of $a^{(B_s\bar{K}^{*})}_{i}$ for the $q^2$-region $[0.1,\,19]\,\gev^2$ are given in \cref{tab:BtoKstarparameters}. 
Information about the uncertainty and correlations is compiled in \cref{app:covarianceObs}. 
\begin{table}[ht!]
 \centering
 \begingroup
  \resizebox{0.47\textwidth}{!}{
   \renewcommand{\arraystretch}{1.95}
 \begin{tabular}{ccccccccccc}
  \hline
  \hline
  \rowcolor{LightBlue}
  $a^{(B_s\bar{K}^{*})}_{1}$ & $a^{(B_s\bar{K}^{*})}_{2}$ &$a^{(B_s\bar{K}^{*})}_{3}$ &$a^{(B_s\bar{K}^{*})}_{4}$ &$a^{(B_s\bar{K}^{*})}_{5}$ &$a^{(B_s\bar{K}^{*})}_{6}$ &$a^{(B_s\bar{K}^{*})}_{7}$ &$a^{(B_s\bar{K}^{*})}_{8}$ &$a^{(B_s\bar{K}^{*})}_{9}$ &$a^{(B_s\bar{K}^{*})}_{10}$ &$a^{(B_s\bar{K}^{*})}_{11}$ \\
  \hline
  45.96 & 7.81 & 0.61 & 9.51 & -12.21 & 47.92 & 0.18 & 1.45 & 1.44 & -15.85 & -0.95   \\
  \hline
  \rowcolor{LightBlue}
  $a^{(B_s\bar{K}^{*})}_{12}$&$a^{(B_s\bar{K}^{*})}_{13}$ & $a^{(B_s\bar{K}^{*})}_{14}$ &$a^{(B_s\bar{K}^{*})}_{15}$ &$a^{(B_s\bar{K}^{*})}_{16}$ &$a^{(B_s\bar{K}^{*})}_{17}$ &$a^{(B_s\bar{K}^{*})}_{18}$ &$a^{(B_s\bar{K}^{*})}_{19}$ &$a^{(B_s\bar{K}^{*})}_{20}$ &$a^{(B_s\bar{K}^{*})}_{21}$ &$a^{(B_s\bar{K}^{*})}_{22}$  \\
  \hline
   -7.52&8.55 & 47.92 & 0.18  & 1.45  & 1.44  & -9.18  & -0.05 & -2.02 & -2.02  & 3.67 \\
 \hline
 \rowcolor{LightBlue}
  $a^{(B_s\bar{K}^{*})}_{23}$&$a^{(B_s\bar{K}^{*})}_{24}$&$a^{(B_s\bar{K}^{*})}_{25}$ & $a^{(B_s\bar{K}^{*})}_{26}$ &$a^{(B_s\bar{K}^{*})}_{27}$ &$a^{(B_s\bar{K}^{*})}_{28}$ &$a^{(B_s\bar{K}^{*})}_{29}$ &$a^{(B_s\bar{K}^{*})}_{30}$ &$a^{(B_s\bar{K}^{*})}_{31}$ &$a^{(B_s\bar{K}^{*})}_{32}$ &$a^{(B_s\bar{K}^{*})}_{33}$   \\
  \hline
  -0.51&-0.51&3.67 & 9.55 & -4.27 & -4.27 & 9.55 & 0.44  & -0.24 & -0.24 & 0.44 \\
  \hline
  \hline
  \end{tabular}
  }
  \endgroup
  \caption{Central values of $a_{i}^{(B_s\bar{K}^{*})}$ 
  in units of $10^{-9}$
  for the $q^2$-region
  $[0.1,\,19]$ \gev$^2$, see \cref{eq:BsKstarPara}. 
  Further details about uncertainties and correlations are provided in \cref{app:covarianceObs}.}
  \label{tab:BtoKstarparameters}
\end{table}

\subsubsection{\texorpdfstring{$B^0\to\mu^+\mu^-$}{B0 to mumu} and scalar operators}\label{sec:Btomumu}

The $B^0\to\mu^+\mu^-$ decay is a clean observable to probe
beyond standard model (BSM) physics in $b\to d\,\mu^+\mu^-$ transitions. 
In the SM, only the operator $\mathcal{O}_{10}$ contributes which yields~\cite{Beneke:2019slt} including updated CKM values~\cite{ParticleDataGroup:2022pth}
\begin{align}\label{eq:SMBtomumu}
    \mathcal{B}(B^0\to\mu^+\mu^-)_{\text{SM}}\,=\,(1.01\pm0.07)\cdot 10^{-10}\,,
\end{align}
in agreement with the experimental value~\cite{LHCb:2021awg}
\begin{align}\label{eq:expBtomumu}
   \mathcal{B}(B^0\to\mu^+\mu^-)_\text{exp}=(1.20 \pm 0.84)\cdot 10^{-10}\,. 
\end{align}
Combining \cref{eq:SMBtomumu,eq:expBtomumu}, we obtain
\begin{align}\label{eq:ratioBtomumuexp}
    \frac{\mathcal{B}(B^0\to\mu^+\mu^-)_\text{exp}}{\mathcal{B}(B^0\to\mu^+\mu^-)_{\text{SM}}}\,=\,1.19\pm 0.84~.
\end{align}

The purely leptonic decay $B^0\to\mu^+\mu^-$  is  sensitive to $\mathcal{O}_{10}^{(\prime)}$, and scalar and pseudoscalar operators,
\begin{align}\label{eq:operatorsSP}
\begin{split}
    \mathcal{O}_S^{(\prime)}\,&=\,(\bar{d}_{L(R)} b_{R(L)})\,(\bar{\mu}\mu)~,\\
    \mathcal{O}_P^{(\prime)}\,&=\,(\bar{d}_{L(R)} b_{R(L)})\,(\bar{\mu}\gamma_5\mu)~.
\end{split}
\end{align}
Including their effects, the branching ratio can be written as~\cite{DeBruyn:2012wk}  
\begin{align}\label{eq:BrBtomumu}
&\frac{\mathcal{B}(B^0\to\mu^+\mu^-)_{\phantom{\text{SM}}}}{\mathcal{B}(B^0\to\mu^+\mu^-)_{\text{SM}}}= \left|\mathcal{P}\right|^2 +|\mathcal{S}|^2 ,
\end{align}
where 
\begin{align}
    \mathcal{P}&=\frac{C_{10}^{\text{SM}}+C_{10^-}}{C_{10}^{\text{SM}}}\,+\,\frac{m_B^2}{2\,m_\mu}\,\left(\frac{1}{m_b+m_d}\right)\,\left(\frac{C_{P^-}}{C_{10}^{\text{SM}}}\right)~,\nonumber\\
    \mathcal{S}&=\frac{m_B^2}{2\,m_\mu}\sqrt{1 - {4m_\mu^2 \over m_B^2}}\,\left(\frac{1}{m_b+m_d}\right)\,\left(\frac{C_{S^-}}{C_{10}^{\text{SM}}}\right)~.
\end{align}
Setting $C_{P^-}$ and $C_{S^-}$ to zero, \cref{eq:ratioBtomumuexp} implies the following ranges for $C_{10^-}$: 
\begin{align} \label{eq:C10mumu}
      -1.8 \lesssim C_{10^-} \lesssim 1.7 \quad\text{or}\quad 6.7 \lesssim C_{10^-}\lesssim 10.1~,
\end{align}
indicating a solution in which  NP is a correction to the SM (the first range), 
and one that involves large cancellations with the SM (the second one), around $C_{10^-}$ near $-2\, {C_{10}^{\text{SM}}}$. 
Future projections for the $300 \,\text{fb}^{-1}$ LHC allow for a reduction of uncertainty by a factor $\sim 6$ \cite{DiCanto:2022icc}. 
Assuming SM central values, this would lead to improved constraints 
\begin{align}\label{eq:C10future}
    -0.3 \lesssim C_{10^-} \! \lesssim 0.3 ~\text{or} ~ 8.1 \lesssim C_{10^-} \! \lesssim 8.6 \,  (\text{HL-LHC}).
\end{align}
NP effects from scalar and pseudoscalar operators \eqref{eq:operatorsSP}
are  more  strongly constrained by data \eqref{eq:ratioBtomumuexp} than $\mathcal{O}_{10}^{(\prime)}$ due to the $m_B/m_\mu$ factor, leading to
\begin{align}
&-0.06 \lesssim C_{P^-} \lesssim 0.05\quad \text{or}\quad 0.2 \lesssim C_{P^-}  \lesssim 0.3~,\label{eq:pscalar}\\
& \hspace{3cm} |C_{S^-}|  \lesssim 0.1 \, ,\label{eq:scalar} 
\end{align}
assuming a single NP coefficient at a time. In this work, we do not consider scalar and pseudoscalar contributions 
in the  global fits, and the $B^0 \to \mu^+ \mu^- $ branching ratio hence depends only on one combination of Wilson coefficients, $C_{10}^\text{SM}+C_{10^-}$. A decomposition of the branching ratio along the lines of \cref{eq:WCC} would result formally in three structures, $\{1, C_{10^-}, C_{10^-}^2\}$, however, all three pre-factors are related, so there is just one coefficient to be varied in the fit.

\subsubsection{\texorpdfstring{$b \to d  \,\gamma$}{b to d gamma}}

The branching ratio of the inclusive decay $\bar B\to X_d\,\gamma$ can be written as~\cite{Hurth:2003dk,Crivellin:2011ba}
\begin{align}\label{eq:btosgamma_th}
    \mathcal{B}(\bar{B}\to X_d\gamma)\,=\,\mathcal{N}\left|\frac{V_{td}^* V_{tb}}{V_{cb}}\right|^2(N+P+P^\prime)~,
\end{align}
where the numerical prefactor $\mathcal{N}= (25.7\pm1.6)\cdot 10^{-4}$ is taken from Ref.~\cite{Hurth:2003dk}. The factor $N=(3.6 \pm 0.6)\cdot 10^{-3}$ encodes non-perturbative corrections~\cite{Gambino:2001ew}, whereas $P$ contains the perturbative SM contributions and the NP effects from $\mathcal{O}_{7,8}$, and $P^\prime$ accounts for NP contributions from $\mathcal{O}_{7,8}^\prime$ operators. 

We updated the SM prediction for the CP-averaged $\bar{B}\to X_d\,\gamma$ branching ratio 
\begin{align}\label{eq:SM_b_to_d_gamma}
\mathcal{B}(\bar{B}\to X_d\,\gamma)_{\text{SM}} &=(16.8\pm 1.7)\cdot10^{-6}\,,
\end{align}
from Ref.~\cite{Misiak:2015xwa}. The main improvement is driven by the updated CKM prefactor $\left|(V_{td}^*V_{tb})/V_{cb}\right|^2\,=\,(41.9 \pm 2.5)\cdot 10^{-3}$~\cite{ParticleDataGroup:2022pth} in \cref{eq:btosgamma_th}. We observe a very good  agreement between the  SM prediction in \cref{eq:SM_b_to_d_gamma} and the extrapolated experimental value from BaBar~\cite{Misiak:2015xwa,BaBar:2010vgu},
\begin{align}\label{eq:Exp_b_to_d_gamma}
    \mathcal{B}(\bar{B}\to X_d\,\gamma)_\text{exp} = \left( 14.1 \pm 5.7 \right) \cdot 10^{-6}\,.
\end{align}
Including NP effects from $C_{7,8}^{(\prime)}$ using Ref.~\cite{Hurth:2003dk}, we find the following parametrization
\begin{align}\label{eq:BXdgamma_para}
   \begin{split}
   \mathcal{B}(\bar{B}\to X_d\,\gamma)&=\sum_{i=1}^9 a^{(\bar{B} X_d)}_{i}\,w^{(\bar{B} X_d)}_{i}~,
   \end{split}
\end{align}
with
\begin{align}
    w^{(\bar{B} X_d)}_{i}\,=\,&\left\lbrace 1,\, C_{7},\, C_{8},\, C_{7}^2,\, C_{8}^2,\, \right.\nonumber\\
    &\left.(C_{7}^\prime)^2,\, (C_{8}^\prime)^2,\,C_7\cdot C_8,\,C_7^\prime\cdot C_8^\prime\right\rbrace \,.
\end{align}
The central values of the factors $a^{(\bar{B} X_d)}_{i}$ are compiled in \cref{tab:BtoXgammaparameters}. These factors suffer from four sources of uncertainties: renormalization scheme dependence of the ratio $m_c/m_b$ ($\sim\,$13 \%), parametric uncertainties ($\sim\,$7 \%), and the perturbative scale $\mu_b$ ($\sim\,$4 \%), which have been included in our analysis.

\begin{table}[ht!]
 \centering
 \begingroup
   \renewcommand{\arraystretch}{1.55}
 \begin{tabular}{ccc}
  \hline
  \hline
  \rowcolor{LightBlue}
  $a_{1}^{(\bar{B} X_d)}$ & $a_{2}^{(\bar{B} X_d)}$ &$a_{3}^{(\bar{B} X_d)}$ \\
  \hline
  $\phm 1.77$ & $-6.17$ & $-0.28$  
  \\
  \hline
  \rowcolor{LightBlue}
  $a_{4}^{(\bar{B} X_d)}=a_{6}^{(\bar{B} X_d)}$ &
  $a_{5}^{(\bar{B} X_d)}=a_{7}^{(\bar{B} X_d)}$ & $a_{8}^{(\bar{B} X_d)}=a_{9}^{(\bar{B} X_d)}$  \\
  \hline
  $\phm 7.66$ & $\phm 0.28$ & $\phm 0.53$  \\
  \hline
  \hline
  \end{tabular}
  \endgroup
  \caption{Central values 
  of $a_{i}^{(\bar{B} X_d)}$ 
  in units of $10^{-5}$, see \cref{eq:BXdgamma_para}. 
}
  \label{tab:BtoXgammaparameters}
\end{table}

Experimental information on the exclusive radiative decay $B\to\rho\,\gamma$ is available~\cite{ParticleDataGroup:2022pth}
\begin{align}
    \mathcal{B}(B^0\to\rho\,\gamma)_{\text{exp}}\,=\,(8.6\pm 1.5)\cdot 10^{-7}~,
\end{align}
however its SM prediction
\begin{align}
    \mathcal{B}(B^0\to\rho\,\gamma)_{\text{SM}}\, \simeq \,(1.0\pm 1.0)\cdot 10^{-6}~,
\end{align}
has sizable uncertainties  from the tensor FF $T_1(0)$. Here, we used  $T_1^{B\to\rho}(0)\,=\,0.24\pm0.12$~\cite{Gubernari:2018wyi}, see
therein for earlier LCSR computations, which are  consistent but have smaller uncertainties.
Confronting the SM to data one obtains the upper limits  $\sqrt{|C_7^{\text{eff}}|^2+|C_7^\prime|^2} \lesssim 0.5 \, (0.3)$ at 90 \% 
C.L.~from $B \to \rho \gamma \, (B \to X_d \gamma)$ branching ratios. 
Since the dependence on the coefficients is the same up to higher order contributions 
and non-perturbative effects in the global fit
we 
only take the inclusive modes which gives stronger constraints into account.
Once the form factor $B \to \rho$ is more accurately known the exclusive decays give stronger constraints as the measurement is more precise.

\section{Global  $b \to d$ Fits} \label{sec:global}

We perform one-dimensional (1D), two-dimensional (2D) and four-dimensional (4D) fits to various combinations and subsets of the BSM-sensitive Wilson coefficients $C_7^{(\prime)},C_8^{(\prime)},C_9^{(\prime)}$ and $C_{10}^{(\prime)}$ in $|\Delta b|=|\Delta d|=1$  transitions. The fit method is described in \cref{sec:fit}. In \cref{sec:res},  the set-up of the analysis and the results of the global $b \to d$ fits are presented. A comparison of our limits on semileptonic four-fermion operators with Drell-Yan and dineutrino constraints is presented in \cref{sec:smeft}.

The aim of  performing  higher dimensional fits is in particular  to work out and demonstrate  the impact and benefits of correlations between different observables and,
 for the case of the 4D fit,  obtain conservative limits for global SMEFT analyses involving semileptonic four-fermion operators.

\subsection{Fit method }\label{sec:fit}

Using the experimental information from $b\to d$ transitions presented in \cref{sec:obs}, we perform global fits to obtain information on the Wilson coefficients $C_i^{(\prime)}$ (collectively denoted here as $\theta$). We work within a frequentist framework based on the approximation of Gaussian likelihood $\mathcal{L}(\theta)=\text{e}^{-\chi^2(\theta)/2}$, resulting in the following $\chi^2$ function
\begin{align}\label{eq:chi2}
\chi^2(\theta)\,=\,-\,2\,\text{ln}\,\mathcal{L}(\theta)\,=\sum_{i,j}^{n_{\text{obs}}}\Delta_i(\theta)\,V_{ij}^{-1}(\theta)\,\Delta_j(\theta)~,
\end{align}
with
\begin{align}
    \Delta_i(\theta)\,&=\,\Delta_i^{(\text{th})}(\theta)-\Delta_i^{(\text{exp})}~,\\
    V_{ij}(\theta)\,&=\,V^{(\text{th})}_{ij}(\theta)+V^{(\text{exp})}_{ij}~.\label{eq:covmatrix}
\end{align}
Here, $n_{\text{obs}}$ represents the number of observables included in the fit ($n_{\text{obs}}=6$), with
\begin{align}
\vec{\Delta}\,=\,\lbrace&\mathcal{B}_1^{(B\pi)},\mathcal{B}_2^{(B\pi)},\mathcal{B}_9^{(B\pi)},\mathcal{B}(B_s^0\to \bar{K}^{*0} \mu^+\mu^-),\nonumber\\
    &\mathcal{B}(B^0\to \mu^+\mu^-),\mathcal{B}(\bar{B}\to X_d\,\gamma)\rbrace\,.
\end{align}
Furthermore, $\Delta_i^{(\text{th})}(\theta)$ is the central value of the theory prediction for the $i$-th observable, $\Delta_i^{(\text{exp})}$ is the central value of the experimental measurement of the same observable, while $V^{(\text{th})}_{ij}(\theta)$ and $V^{(\text{exp})}_{ij}$ are the theoretical and experimental covariance matrices, respectively. 
We provide details on the computation of these matrices in \cref{app:covarianceObs}. To estimate the central values of the $\theta$ parameters, we use the maximum likelihood (ML) method. 
By construction of $\mathcal{L}(\theta)$, the ML estimators $\hat{\theta}$ correspond to the best-fit points obtained by minimizing the $\chi^2(\theta)$ function, that is $\partial\,\chi^2/\partial\,\theta_i|_{\hat{\theta}}=0$ with $i=1,...,n_{\text{par}}$ and $n_{\text{par}}$ being the number of parameters (up to 8 parameters in the most general scenario). In practise, we use \ALmigrad from the \python package \iminuit~\cite{hans_dembinski_2022_6330776} to conduct the numerical minimization. 
We scan for the viable parametric space of $\theta$ by initializing the \ALmigrad algorithm to different initial values of $\theta$. 
In some cases, we find multiple solutions with similar $\chi^2_\text{min}\equiv\chi^2(\hat{\theta})$, we always choose the one with a smaller $\chi^2$ value. In that sense, future measurements of $b \to d\, \,\mu^+ \mu^-$ modes are necessary to exclude other solutions with larger $\chi^2$ values.  

The confidence regions for $l=1,2,...$ sigmas are worked out via $\Delta\chi^2(\theta)\leq \eta\,(l,n_{\text{par}})$ where $\Delta\chi^2(\theta)=\chi^2(\theta)-\chi^2_{\text{min}}$ and $\eta\,(l,n_{\text{par}})$ is the value where the $\chi^2$ cumulative distribution (CDF) function reaches the probability associated with $l$ sigmas, {\it e.g.} $\eta\,(l,1)=l^2$, $\eta\,(l,2)=(2.30,6.18,...)$, etc. In practise, the confidence intervals are computed using \ALminos algorithm from \iminuit~\cite{hans_dembinski_2022_6330776}.

With the explicit dependence of the $\chi^2$ function given by \cref{eq:chi2}, we compute the errors, $\sigma_{\hat{\theta}_i}=\sqrt{(V_{\hat{\theta}})_{ii}}$, and the correlations among the fit parameters, $\rho_{\hat{\theta}_i\hat{\theta}_j}\equiv(V_{\hat{\theta}})_{ij}/\sqrt{(V_{\hat{\theta}})_{ii}(V_{\hat{\theta}})_{jj}}$, which are given by the matrix
\begin{align}
    (V_{\hat{\theta}}^{-1})_{ij}\,=\,\left.\frac{1}{2}\frac{\partial^2\chi^2}{\partial\theta_i\partial\theta_j}\right|_{\hat{\theta}}~.
\end{align}
These matrices are provided as an output by \ALmigrad when performing the minimization routine. 

To determine the level of statistical significance between a given hypothesis $H$ and the data, we also determine the $p-$value of the respective fit,
\begin{align}\label{eq:pvalue}
    p_H=1-\text{CDF}(\chi^2_{\text{min}};n_{\text{dof}})~,
\end{align}
where $n_\text{dof}=n_\text{obs}-n_\text{par}$ is the number of degrees of freedom. 
A $p-$value of less than 0.05 indicates strong evidence against the assumed hypothesis, as there is less than a 5\% probability that $H$ is correct. Finally, to compare the description offered by two hypotheses $H$ and SM, we provide their relative pull (in units of standard deviations $\sigma$)
\begin{align}\label{eq:pull}
    \text{Pull}_{H}\,=\,\sqrt{2}\,\text{Erf}^{-1}\left[\text{CDF}(\Delta\chi^2_{H};n_{H})\right]~,
\end{align}
where $\text{Erf}$ is the error function, $\Delta\chi^2_{H}=\chi^2_{\text{SM},\text{min}}-\chi^2_{H,\text{min}}$, and $n_H$ is the number of parameters in $H$.

\subsection{Results for Wilson coefficient fits} \label{sec:res}

In the following, we present the fit results for 1D, 2D, and 4D scenarios floating different combinations of NP Wilson coefficients. 
The main results of the fits are given in \cref{tab:fitres_1d,tab:fitres_2d,tab:fitres_4d}. Constraints on NP Wilson coefficients $C_i,C_i^\prime$ are given at the $b$-quark mass scale. The SM hypothesis gives a good fit to $b\to d\,\mu^+\mu^-$ and $b\to d\,\gamma$ data ($n_\text{obs}\,=\,6$), resulting in $\chi^2_\text{min}\,=\,3.77$ and a $p-$value of 71 \%.

\begingroup
\renewcommand*{\arraystretch}{1.6}
\begin{table*}[ht!]
 \centering
 \begin{tabular}{c|c|c|c|c|c|c|c}
  \hline
  \hline
  \rowcolor{LightBlue}
  scenario  & fit parameter & best fit &$1\sigma$  &$2\sigma$  &$\chi^2_{H_i,\,\text{min}}$ & Pull$_{H_i}$ & $p$-value (\%) \\ 
  \hline
  \hline
  $H_{1}$ &                                                   $C_7$ &      0.01 &   [-0.07, 0.11] &  [-0.15, 0.25] &                       3.74 &                 0.15 &             58 \\
  $H_{2}$ &                                                   $C_8$ &      0.04 &   [-0.88, 1.44] &  [-1.51, 2.27] &                       3.76 &                 0.04 &             58 \\
  $H_{3}$ &                                                   $C_9$ &     -1.37 &  [-2.97, -0.47] &  [-7.65, 0.26] &                       1.12 &                 1.63 &             95 \\
  $H_{4}$ &                                                $C_{10}$ &      0.96 &     [0.3, 1.75] &  [-0.29, 2.92] &                       1.51 &                 1.50 &             91 \\
  $H_{5}$ &                                            $C_7^\prime$ &     -0.02 &   [-0.18, 0.16] &   [-0.31, 0.3] &                       3.75 &                 0.11 &             58 \\
  $H_{6}$ &                                            $C_8^\prime$ &     -0.04 &   [-1.16, 1.13] &  [-1.86, 1.85] &                       3.76 &                 0.03 &             58 \\
  $H_{7}$ &                                            $C_9^\prime$ &     -0.21 &   [-0.91, 0.47] &  [-1.63, 1.15] &                       3.67 &                 0.32 &             59 \\
  $H_{8}$ &                                         $C_{10}^\prime$ &      0.22 &    [-0.37, 0.8] &  [-0.98, 1.38] &                       3.63 &                 0.37 &             60 \\
  $H_{9}$ &                                           $C_9=+C_{10}$ &      0.19 &   [-0.57, 1.02] &  [-1.24, 1.79] &                       3.71 &                 0.24 &             59 \\
 $H_{10}$ &                                           $C_9=-C_{10}$ &     -0.53 &  [-0.89, -0.19] &  [-1.29, 0.14] &                       1.27 &                 1.58 &             93 \\
 $H_{11}$ &                             $C_9^\prime=+C_{10}^\prime$ &      0.10 &   [-0.68, 0.86] &  [-1.41, 1.53] &                       3.75 &                 0.13 &             58 \\
 $H_{12}$ &                             $C_9^\prime=-C_{10}^\prime$ &     -0.13 &   [-0.46, 0.22] &   [-0.8, 0.57] &                       3.63 &                 0.37 &             60 \\
 $H_{13}$ &                                $C_9=-C_9^\prime$ &     -1.74 &  [-3.26, -0.27] &  [-4.04, 0.44] &                       1.96 &                 1.34 &             85 \\
 $H_{14}$ &                                $C_9=+C_9^\prime$ &     -0.55 &  [-1.29, -0.07] &  [-4.13, 0.34] &                       2.42 &                 1.16 &             78 \\
 $H_{15}$ &  $C_9=-C_{10}=-C_9^\prime=-C_{10}^\prime$ &     -0.58 &   [-1.06, -0.2] &  [-4.04, 0.12] &                       1.17 &                 1.61 &             94 \\
 $H_{16}$ &  $C_9=-C_{10}=+C_9^\prime=-C_{10}^\prime$ &     -0.24 &  [-0.46, -0.04] &   [-0.7, 0.16] &                       2.35 &                 1.19 &             79 \\
  \hline  
  \hline
  \end{tabular}
  \caption{Fit results for Wilson coefficients in the 1D scenarios ($H_{1,...,16}$) at the scale $\mu_b$. Best-fit values and $1\sigma$ ($2\sigma$) uncertainties are displayed in the third and fourth (fifth) columns. The indicators of the goodness-of-fit are provided in the last columns: $\chi^2$ function evaluated at the best-fit point; the Pull$_{H_i}$ in units of standard deviation, as in \cref{eq:pull}; and the $p-$value $p_{H}$, as in \cref{eq:pvalue}. The SM has $\chi^2_\text{SM}=3.77$, with a $p-$value of 71\%.
  }
  \label{tab:fitres_1d}
\end{table*}
\endgroup

\subsubsection{One-dimensional fits}

We start with 1D scenarios ($H_{1,...,16}$), which are
compiled in \cref{tab:fitres_1d}. Here, the first column displays the name of the scenario, while the second column indicates the fitted Wilson coefficient and the correlations assumed between other Wilson coefficients, as in  $H_{9,...,16}$. The best-fit values with their 1$\sigma$ and 2$\sigma$ confidence intervals are given in the third and fourth columns, respectively. The last columns show the indicators of the goodness-of-fit: best-fit $\chi^2$ value, the pull from the SM hypothesis, see \cref{eq:pull}, and the $p-$value of the fit, see \cref{eq:pvalue}. 

Among the  1D scenarios with only one Wilson coefficient switched on at a time, $H_{1,...,8}$, the most favored one is $H_3$ with NP in $C_9$ only, with a pull of $1.63\,\sigma$ and a $p-$value of 95\%.  
The second most favored scenario is  $H_4$ with a pull of $1.50\,\sigma$ and a $p-$value of 91\%, where NP only enters in $C_{10}$. 
One observes that the global fit selects the solution with the smaller NP effect from the ranges given by $B^0 \to \mu^+ \mu^-$ alone, see \cref{eq:C10mumu}.

\begingroup
\renewcommand*{\arraystretch}{1.6}
\begin{table*}[ht!]
 \centering
 \resizebox{0.95\textwidth}{!}{
 \begin{tabular}{c|c|c|c|c|c|c|c}
  \hline
  \hline
  \rowcolor{LightBlue}
  scen.  & fit parameters  & best fit & $1\sigma$ & $2\sigma$ & $\chi^2_{H_i,\,\text{min}}$ & Pull$_{H_i}$  & $p$-v. (\%) \\ 
  \hline
  \hline
 $H_{17}$ &                                ($C_7$,\,$C_8$) &   (0.02, -0.13) &   ([-0.08, 0.21], [-1.57, 1.46]) &  ([-0.16, 0.43], [-2.37, 2.41]) &                       3.74 &                 0.02 &             44 \\
 $H_{18}$ &                                ($C_7$,\,$C_9$) &   (0.05, -1.45) &   ([-0.04, 0.19], [-3.0, -0.54]) &  ([-0.13, 0.94], [-9.65, 0.22]) &                       0.86 &                 1.19 &             93 \\
 $H_{19}$ &                             ($C_7$,\,$C_{10}$) &    (0.04, 1.02) &    ([-0.05, 0.17], [0.34, 1.87]) &  ([-0.13, 0.73], [-0.26, 3.32]) &                       1.33 &                 1.05 &             85 \\
 $H_{20}$ &                         ($C_7$,\,$C_7^\prime$) &   (0.01, -0.02) &   ([-0.07, 0.12], [-0.21, 0.17]) &  ([-0.15, 0.28], [-0.36, 0.34]) &                       3.73 &                 0.02 &             44 \\
 $H_{21}$ &                         ($C_7$,\,$C_9^\prime$) &   (0.02, -0.23) &   ([-0.07, 0.12], [-0.92, 0.46]) &  ([-0.15, 0.26], [-1.64, 1.15]) &                       3.63 &                 0.08 &             45 \\
 $H_{22}$ &                      ($C_7$,\,$C_{10}^\prime$) &    (0.02, 0.23) &   ([-0.07, 0.12], [-0.36, 0.81]) &   ([-0.15, 0.26], [-0.97, 1.4]) &                       3.60 &                 0.10 &             46 \\
 $H_{23}$ &                             ($C_9$,\,$C_{10}$) &   (-1.67, 8.55) &    ([-7.43, 0.65], [6.48, 9.37]) &  ([-9.13, 1.86], [-1.42, 9.85]) &                       1.00 &                 1.15 &             91 \\
 $H_{24}$ &                         ($C_7^\prime$,\,$C_9$) &    (0.05, -1.4) &  ([-0.17, 0.23], [-2.95, -0.49]) &  ([-0.35, 0.36], [-7.64, 0.26]) &                       1.08 &                 1.12 &             89 \\
 $H_{25}$ &                         ($C_9$,\,$C_9^\prime$) &   (-2.22, 1.18) &  ([-6.55, -0.63], [-2.99, 2.89]) &  ([-7.58, 0.23], [-3.92, 3.81]) &                       0.76 &                 1.22 &             94 \\
 $H_{26}$ &                      ($C_9$,\,$C_{10}^\prime$) &  (-1.79, -0.35) &  ([-6.59, -0.57], [-1.19, 0.36]) &   ([-7.61, 0.27], [-1.8, 1.05]) &                       0.88 &                 1.18 &             92 \\
 $H_{27}$ &                      ($C_7^\prime$,\,$C_{10}$) &    (0.04, 0.99) &    ([-0.16, 0.22], [0.31, 1.84]) &   ([-0.3, 0.35], [-0.29, 3.25]) &                       1.48 &                 1.00 &             83 \\
 $H_{28}$ &                      ($C_9^\prime$,\,$C_{10}$) &    (0.21, 7.34) &    ([-0.58, 0.99], [6.29, 8.09]) &   ([-1.39, 1.79], [-0.3, 8.72]) &                       1.35 &                 1.04 &             85 \\
 $H_{29}$ &                   ($C_{10}$,\,$C_{10}^\prime$) &   (7.45, -0.01) &    ([6.53, 8.13], [-0.79, 0.97]) &  ([-0.30, 8.73], [-4.54, 4.49]) &                       1.42 &                 1.02 &             84 \\
 $H_{30}$ &                  ($C_7^\prime$,\,$C_9^\prime$) &   (0.02, -0.26) &   ([-0.18, 0.21], [-1.07, 0.57]) &  ([-0.32, 0.34], [-1.88, 1.36]) &                       3.66 &                 0.07 &             45 \\
 $H_{31}$ &               ($C_7^\prime$,\,$C_{10}^\prime$) &     (0.0, 0.22) &   ([-0.17, 0.18], [-0.41, 0.84]) &  ([-0.31, 0.32], [-1.04, 1.46]) &                       3.63 &                 0.08 &             45 \\
 $H_{32}$ &               ($C_9^\prime$,\,$C_{10}^\prime$) &   (-0.08, 0.17) &   ([-1.07, 0.83], [-0.65, 0.99]) &  ([-2.04, 1.63], [-1.39, 1.74]) &                       3.62 &                 0.09 &             45 \\
 $H_{33}$ &  $( C_9=-C_9^\prime,\,C_{10}=+C_{10}^\prime )$ &   (-1.73, 0.44) &   ([-3.34, -0.19], [0.04, 0.95]) &   ([-4.1, 0.51], [-0.34, 4.52]) &                       0.77 &                 1.22 &             94 \\
 $H_{34}$ &  $( C_9=-C_9^\prime,\,C_{10}=-C_{10}^\prime )$ &   (-1.73, 0.01) &   ([-3.65, 0.15], [-0.45, 0.91]) &   ([-4.6, 1.05], [-0.84, 5.06]) &                       1.96 &                 0.83 &             74 \\
 $H_{35}$ &  $( C_9=+C_9^\prime,\,C_{10}=+C_{10}^\prime )$ &     (0.6, 2.18) &    ([0.26, 0.89], [-0.58, 4.77]) &   ([-4.95, 1.19], [-0.92, 5.1]) &                       2.15 &                 0.76 &             70 \\
 $H_{36}$ &  $( C_9=-C_{10},\,C_9^\prime=+C_{10}^\prime )$ &   (-0.58, 0.57) &   ([-3.11, -0.2], [-1.37, 3.38]) &  ([-8.03, 0.13], [-3.14, 4.05]) &                       1.17 &                 1.10 &             88 \\
 $H_{37}$ &  $( C_9=-C_{10},\,C_9^\prime=-C_{10}^\prime )$ &    (-0.6, 0.15) &  ([-1.07, -0.21], [-0.27, 0.65]) &  ([-1.86, 0.15], [-0.66, 1.47]) &                       1.15 &                 1.10 &             88 \\
  \hline
  \hline
  \end{tabular}
  }
  \caption{Fit results for the  2D scenarios with similar description as \cref{tab:fitres_1d}.
  }
  \label{tab:fitres_2d}
\end{table*}
\endgroup

\begin{figure*}[ht!]
    \centering
    \includegraphics[width=\columnwidth]{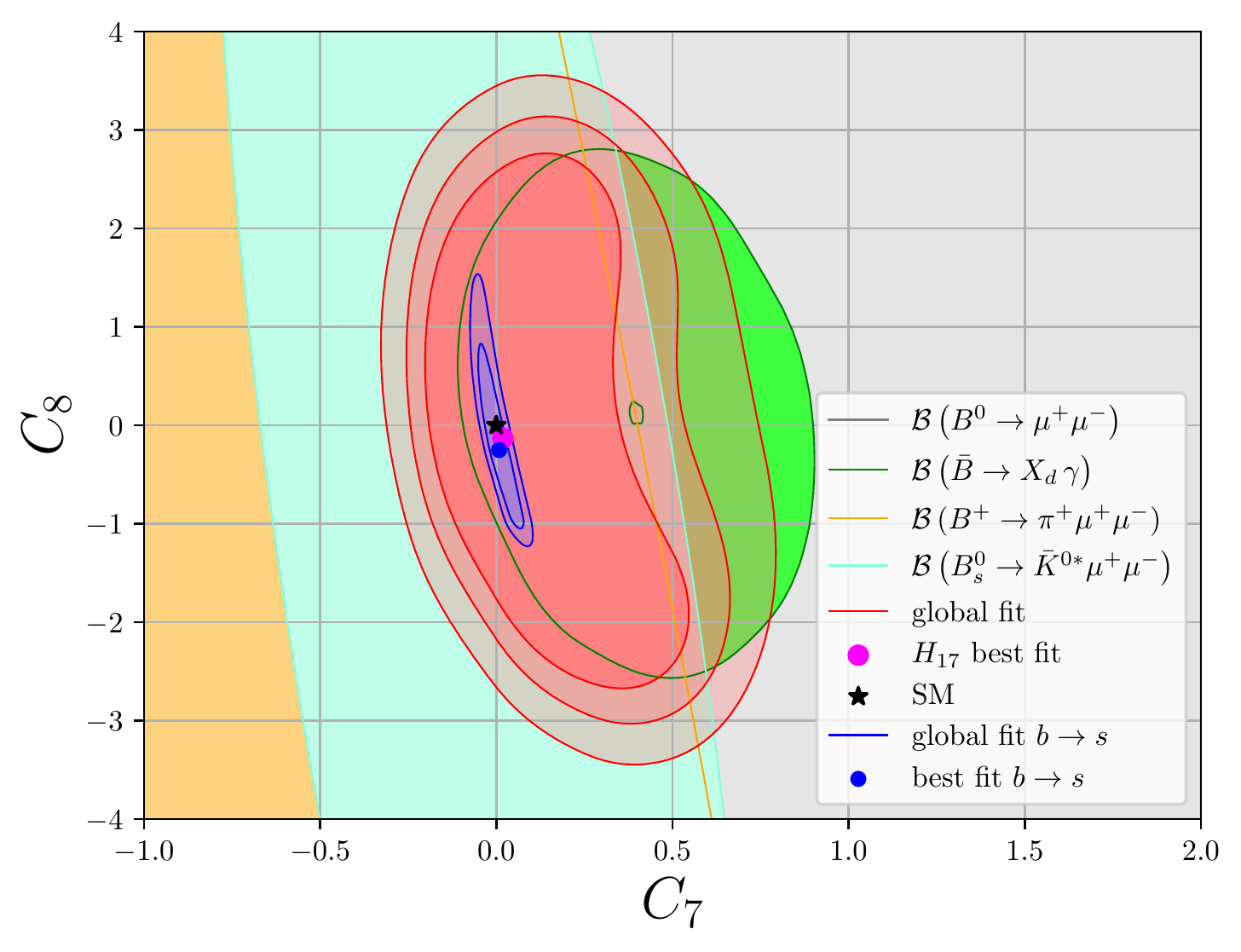}
    \includegraphics[width=\columnwidth]{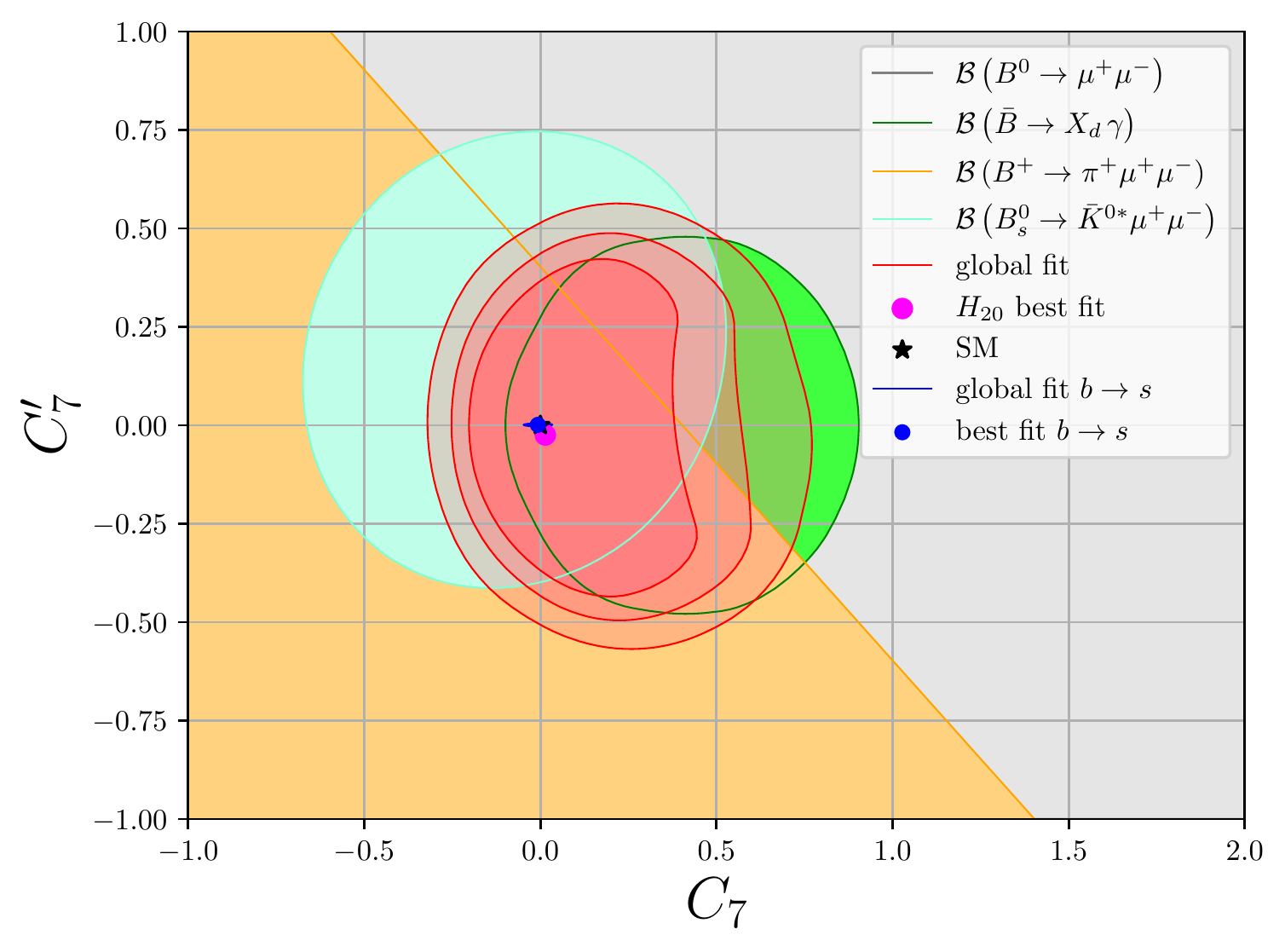}
    \caption{
   Results of  the 2D fits to  Wilson coefficients $C_7$,$C_8$ (scenario $H_{17}$,  left panel) and $C_7$,$C_7^\prime$ (scenario $H_{20}$, right panel), see \cref{tab:fitres_2d}. We show the 1$\sigma$ allowed regions of the individual observables and the combined 1, 2, and $3\sigma$ regions (red). The black star at the origin represents the SM. Results of the global $b \to s$ fit, see \cref{app:btosfit}, are superimposed (blue) and serve as a data-driven prediction assuming minimal flavor violation (MFV). Deviations between the blue area and the SM are a result of the present discrepancies in $b\to s$ processes with the SM. See text for details.
    }
    \label{fig:2dfit_contours_C78}
\end{figure*}

\begin{figure*}[ht!]
    \centering
    \includegraphics[width=\columnwidth]{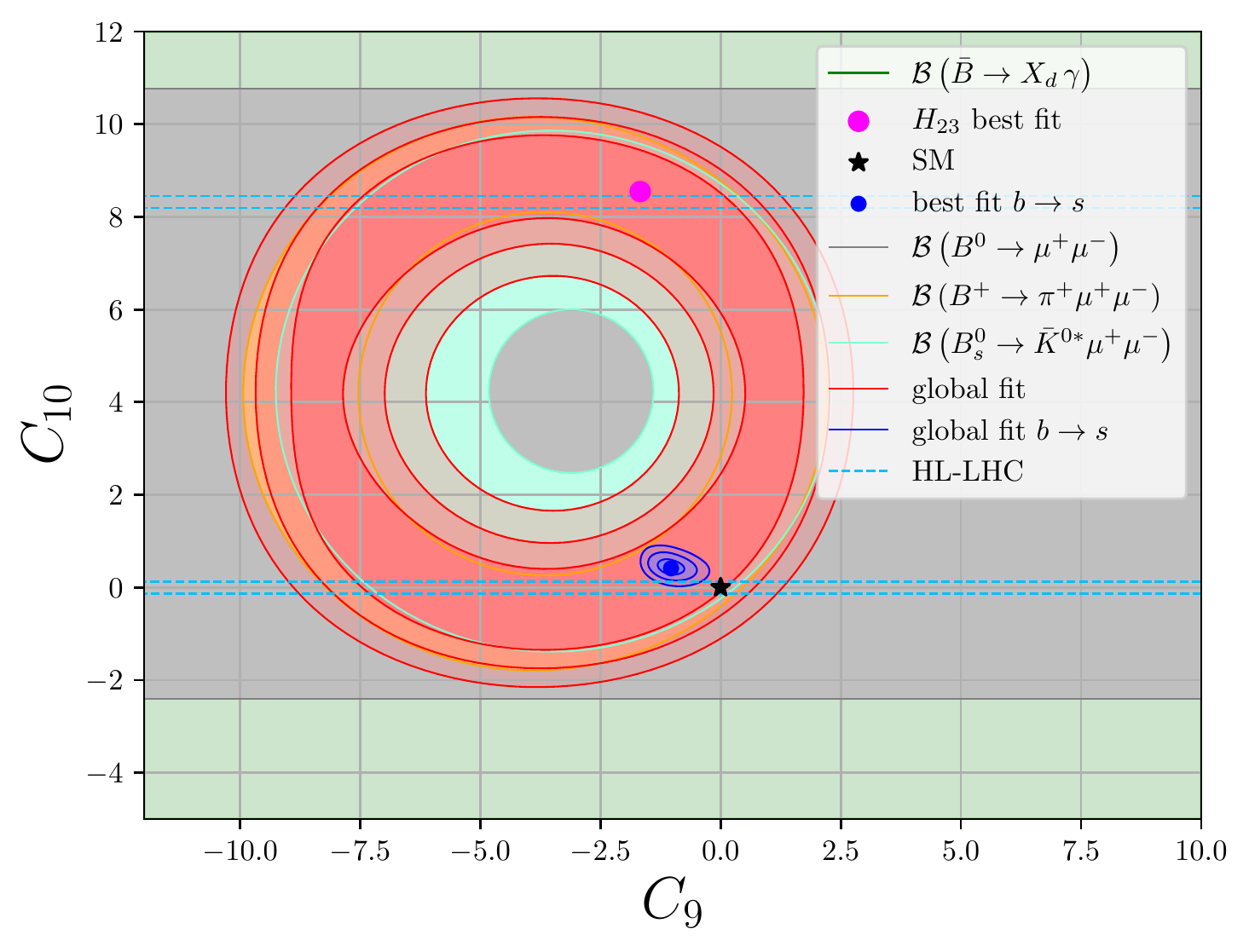}
    \includegraphics[width=\columnwidth]{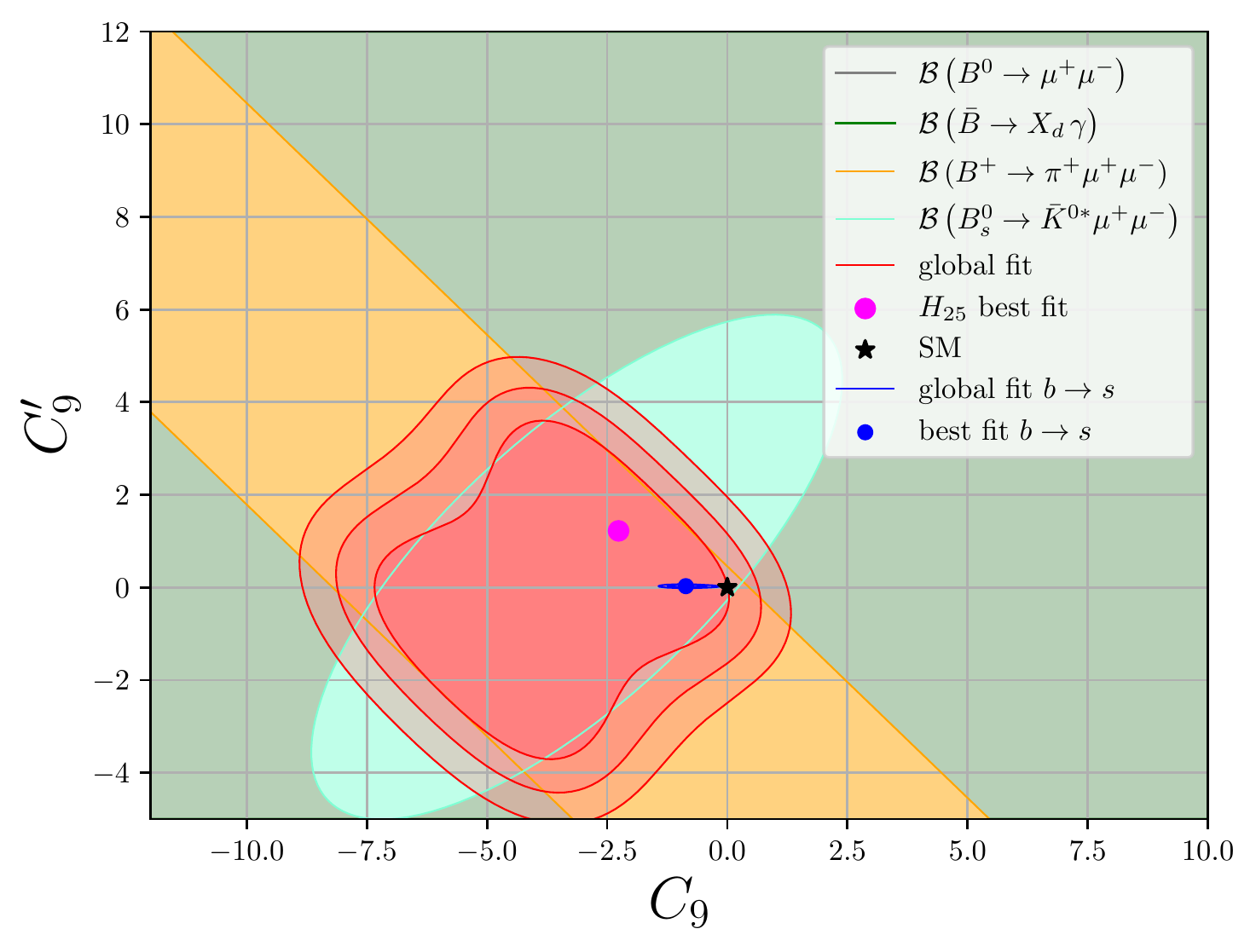}
    \includegraphics[width=\columnwidth]{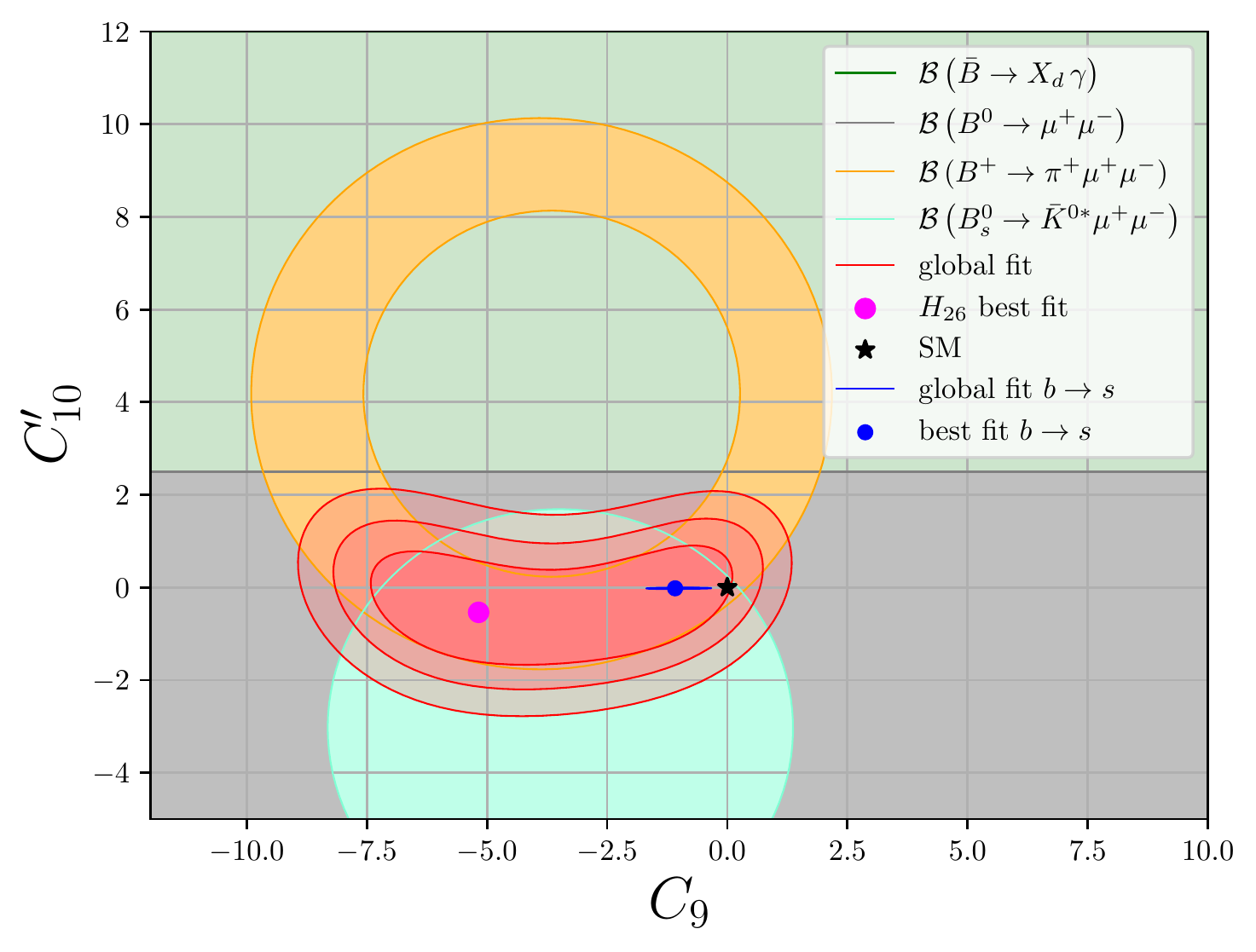}
    \includegraphics[width=\columnwidth]{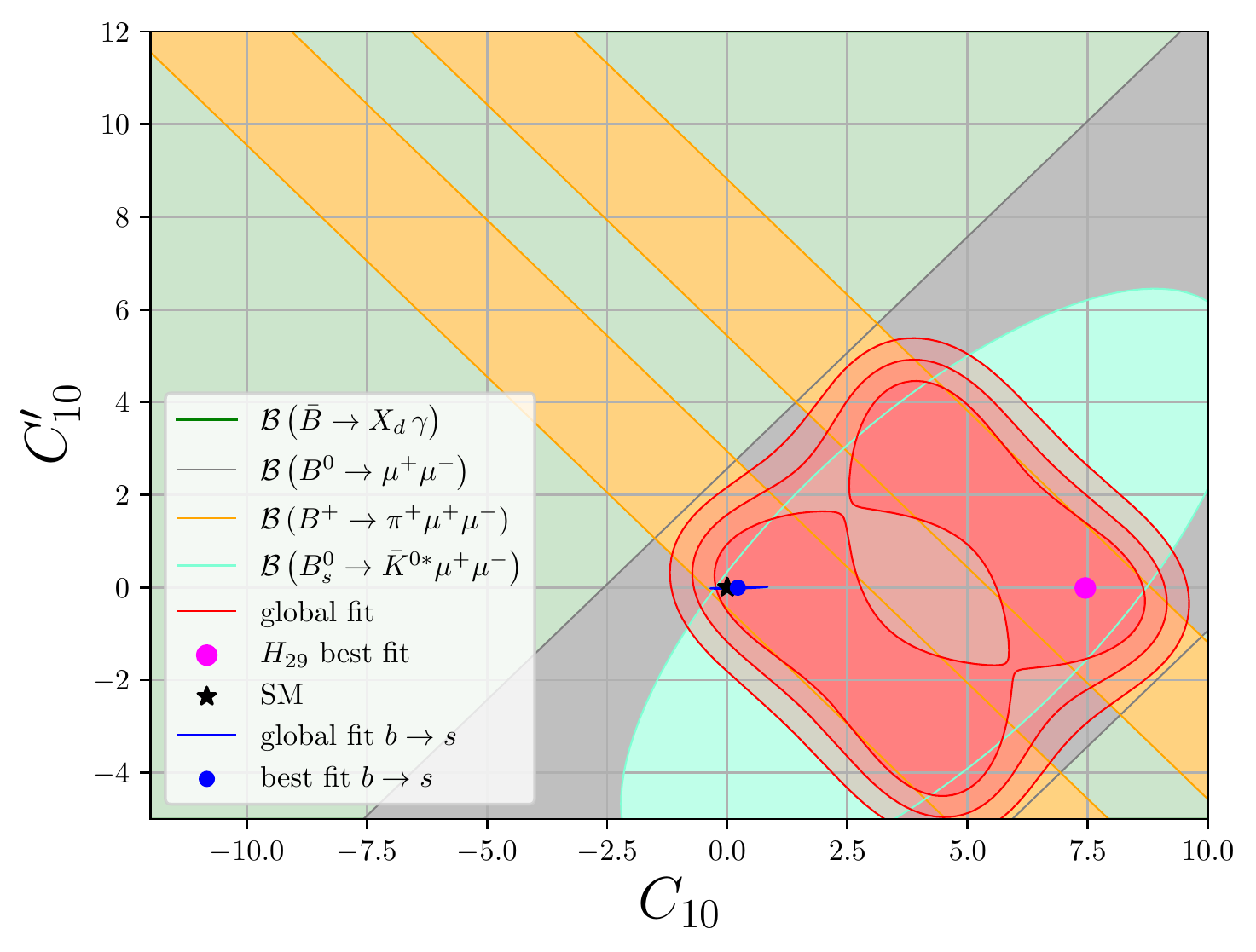}
    \includegraphics[width=\columnwidth]{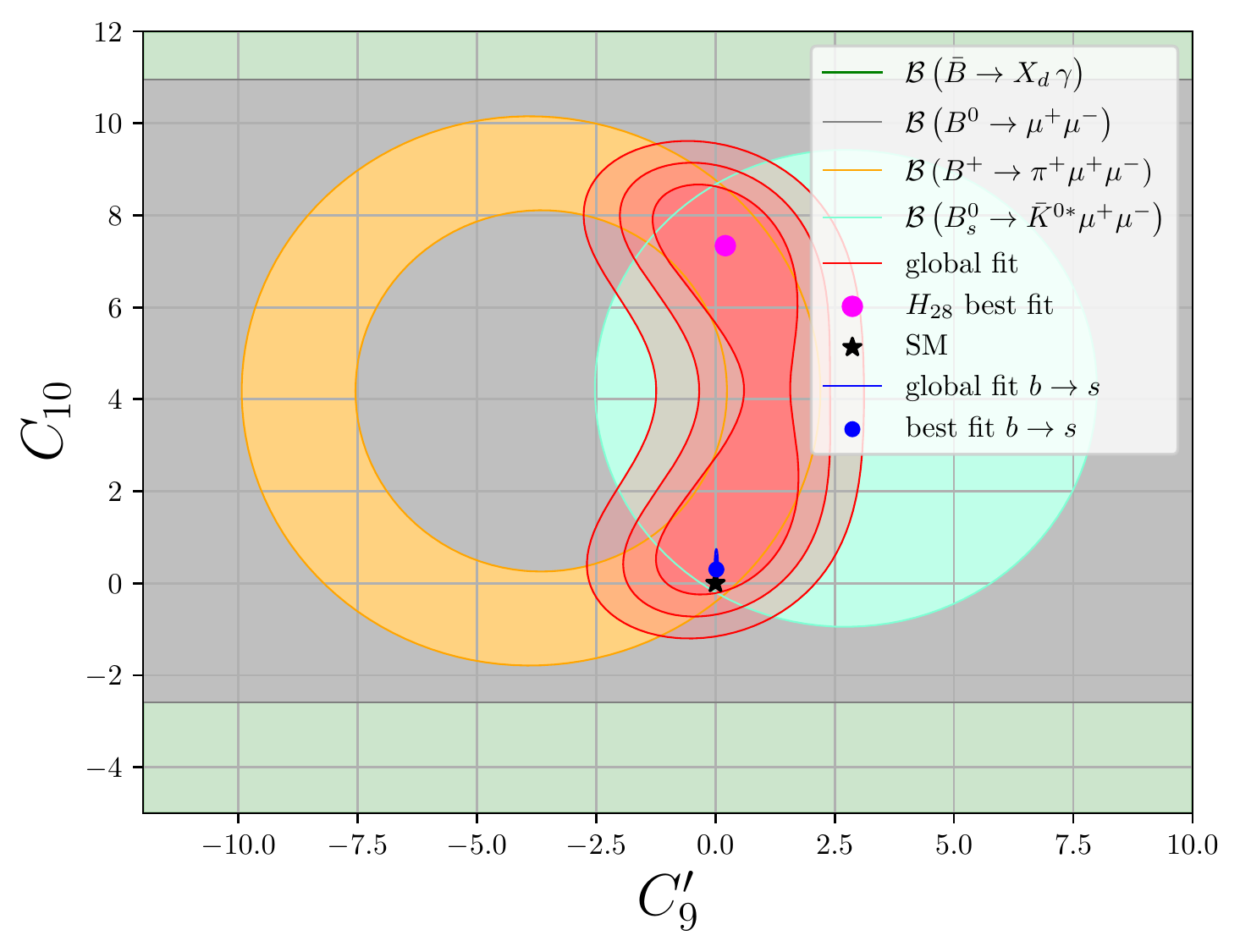}
    \includegraphics[width=\columnwidth]{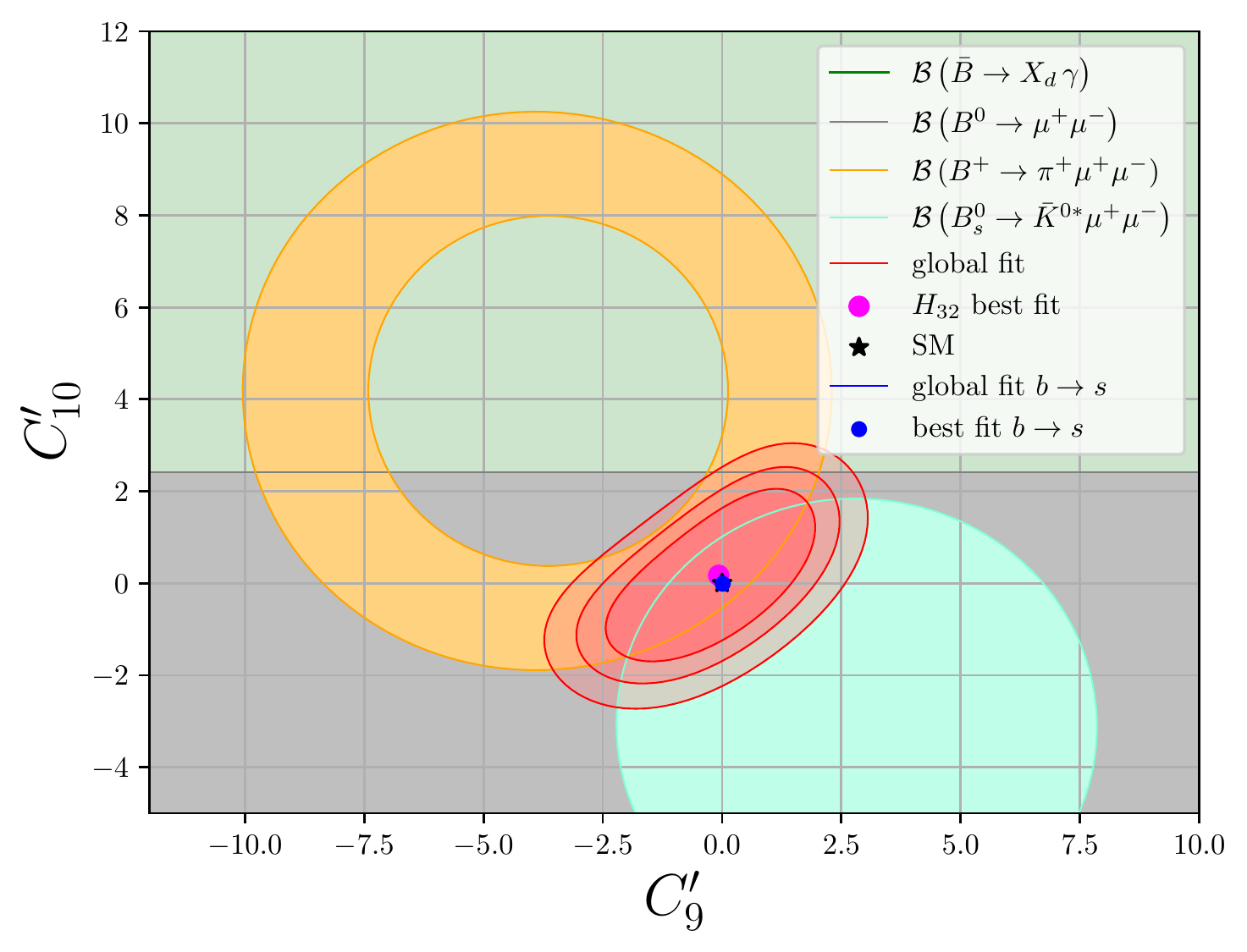} 
    \caption{
Results of  the 2D fits to semileptonic four-fermion operators, see \cref{tab:fitres_2d}. Similar description as in \cref{fig:2dfit_contours_C78}.
    }
    \label{fig:2dfit_contours}
\end{figure*}

Furthermore, we perform global fits in scenarios relating two Wilson coefficients ($H_{9, ...,14}$) in which $SU(2)_L$-invariance is preserved in the lepton sector or those with a preference for leptonic vector-couplings. We find that the scenario $H_{10}$ with left-handed quarks and left-handed leptons, $C_9=-C_{10}$, is preferred by data with a pull of $1.58\,\sigma$ and a $p-$value 93\%. 
For comparison, benchmark $H_{9}$, that is,
left-handed quarks and right-handed leptons $C_9=C_{10}$, only has a pull of $0.24\,\sigma$ and a $p-$value 59\%, worse than the SM one. 
These findings are similar to the ones  in $b \to s$ transitions where a good fit is obtained for $C_9^{(b \to s)}=-C_{10}^{(b \to s)}=-0.50 \pm 0.13$, see \cref{app:btosfit}. 
We explore additional correlations among the primed coefficients $C_{9,10}^{\prime}$ exploiting $C_9^\prime=\pm \,C_{10}^\prime$, see scenarios $H_{11}$ and $H_{12}$ with $p-$values worse than the SM one. 
In addition, we consider $C_9= \pm\, C_9^\prime$, which gives a pull of $1.34\,\sigma$ and a $p-$value of 85\% for $C_9= -C_9^\prime$, 
while a $p$-value of $78\,$\% is obtained for $C_9= +C_9^\prime$. 
Taking into account the previous observations, we perform a fit in scenarios involving four NP coefficients, $H_{15}$ and $H_{16}$, where the most preferred scenario is $H_{15}$ with a pull of $1.61\,\sigma$ and a $p-$value 94\%.

We observe that data shows a clear preference to include NP via $C_9$, following a similar pattern as in global fits to  $b\to s\,\mu^+\mu^-$ data, 
where $C_9^{(b \to s)}=-0.85\pm0.21$ gives a good fit, see \cref{app:btosfit}.
Therefore, future data is very welcome to confirm or refute this pattern in other flavor sectors, 
such as $b \to d $ transitions.

To summarize, while there is consistency with the SM, we also observe room for sizable, order one  NP contributions to Wilson coefficients, significantly larger than corresponding ranges in the $b \to s$ global fit. We note also that the likelihood is quite flat, making uncertainty ranges more sensible than best-fit points. These features  continue to hold in 2D and 4D fits discussed next.

\subsubsection{Two-dimensional fits}

Here we present the results for the 2D fits which are compiled in \cref{tab:fitres_2d}. We exploit scenarios composed out of two coefficients from $C_7,C_7^\prime,C_8,C_9,C_9^\prime,C_{10},C_{10}^\prime$ in $H_{17,...,32}$. 
The presently most favored scenario is $H_{25}$ with a pull of $1.22\,\sigma$ and a $p-$value of 94\%, fitting simultaneously $C_9$ and $C_9^\prime$. We observe again a similar pattern as in the 1D fits, that, if $C_9$ is included in the fit, $p-$values are large, $\sim 90\%$, {\it i.e. $H_{18,23,24,25,26}$}. 
In addition, we perform fits in scenarios where NP enters in a correlated way in four Wilson coefficients, $H_{33,...,37}$.  Among these, scenario $H_{33}$, with $C_9=-C_9^\prime$ and $C_{10}=+C_{10}^\prime$,  features the highest pull of $1.22\,\sigma$ and a large $p-$value of $94\%$. 
Some of these benchmark models, $H_{34,35,37}$ satisfy the relation $C_{10}^\prime/C_{10} = C_9^\prime/C_9$,
which arises from tree-level exchange of a $Z^\prime$-boson~\cite{Descotes-Genon:2015uva}.

In \cref{fig:2dfit_contours_C78}, we display two selected 2D contours of scenarios with dipole coefficients, $H_{17}$ ($C_7$ and $C_8$) and $H_{20}$ ($C_7$ and $C_7^\prime$). One observes an excellent complementarity between $\bar{B}\to X_d \gamma$, $B^+\to\pi^+\mu^+\mu^-$, and $B^0_s\to \bar{K}^{0*}\mu^+\mu^-$ observables. This leads to improved limits on $C_7^{(\prime)}$ compared to previous works \cite{Crivellin:2011ba} (assuming real-valued Wilson coefficients).

\begingroup
\renewcommand*{\arraystretch}{1.6}
\begin{table*}[ht!]
 \centering
 \resizebox{0.95\textwidth}{!}{
 \begin{tabular}{c|c|c|c|c|c|c}
  \hline
  \hline
    \rowcolor{LightBlue}
  scen.  & fit parameter & $1\sigma$ & $2\sigma$  & $\chi^2_{H_i,\,\text{min}}$ & Pull$_{H_i}$  & $p$-v. $(\%)$\\ 
  \hline
  \hline
 $H_{38}$ &  ($C_9$,\,$C_9^\prime$,\,$C_{10}$,\,$C_{10}^\prime$) &  ([-7.96, 0.84], [-4.42, 4.34], [-0.89, 9.31], [-5.12, 5.08]) &  ([-9.06, 1.83], [-5.45, 5.36], [-1.40, 26.4], [-21.8, 5.59]) &                       0.68 &                 0.61 &             71 \\
 $H_{39}$ &                   ($C_7$,\,$C_8$,\,$C_9$,\,$C_{10}$) &  ([-0.04, 0.85], [-2.29, 2.53], [-9.7, 0.26], [-1.06, 1.97]) &    ([-0.13, 2.09], [-2.74, 2.98], [-11.05, 1.77], [-1.55, 9.94]) &                       0.53 &                 0.65 &             76 \\
  \hline
  \hline
  \end{tabular}}
  \caption{Fit results for the  4D scenarios ($H_{38,39}$). Further notation  as in \cref{tab:fitres_1d,tab:fitres_2d}. 
  }
  \label{tab:fitres_4d}
\end{table*}
\endgroup

In \cref{fig:2dfit_contours}, we display contours from fits to semileptonic four-fermion operators. We observe that the complementarity between the observables is currently not as good as for dipole coefficients, leading to weaker limits on $C_9$ and $C_{10}$, 
see the uppermost left plot of \cref{fig:2dfit_contours}, 
displaying contours in $H_{23}$, $C_9$ versus $C_{10}$. 
The branching ratios of $B^+\to\pi^+\mu^+\mu^-$, and $B^0_s\to \bar{K}^{0*}\mu^+\mu^-$ cooperate to reduce the thickness of the annulus (red area) but do not lift the degeneracy between $C_9$ and $C_{10}$. 
The branching ratio of $B^0\to\mu^+\mu^-$ can in principle do so due to its dependence on $C_{10}^{(\prime)}$ only, however,  the present precision is insufficient. 
In the future, the HL-LHC run can improve this significantly, see \cref{eq:C10future} for the projected sensitivity, up to a discrete ambiguity.
This  is illustrated by the dashed blue horizontal lines in the uppermost left plot of \cref{fig:2dfit_contours}. 
Note also that due to the flat likelihood along the ring (red area) the best-fit point (magenta) is only shown for completeness but has little statistical preference over other points in this flat direction. The projections in \cref{fig:2dfit_contours} also make visible discrete ambiguities, for instance the two yellow bands in $C_{10}, C_{10}^\prime$ (mid right-handed plot).

To remove the ambiguities additional complementary observables of $b \to d \,\ell \ell$ transitions are necessary. As well known from the early days of the $b \to s \,\mu \mu $ fits, angular observables break the degeneracy in the branching ratio based fits and significantly improve the information on the NP couplings \cite{Ali:2002jg,Bobeth:2010wg}.
The most simple one is the forward-backward asymmetry $A_{\text{FB}}^\ell$ in the lepton angle,
defined as
\begin{align} \label{eq:AFB}
A_{\text{FB}}^\ell \propto \int_0^1 \!\!\text{d} \cos \theta_\ell  \frac{\text{d}^2 \Gamma}{\text{d} q^2 \text{d} \cos \theta_\ell} -\int_{-1}^0 \!\! \text{d} \cos \theta_\ell  \frac{\text{d}^2 \Gamma}{\text{d} q^2 \text{d}\cos \theta_\ell}
\end{align}
and $\theta_\ell$ is  the angle between the negatively charged lepton
and the decaying $b$-hadron  in the dilepton center of mass system. 
In $1/2 \to 1/2  \,\ell^+ \ell^- $ baryon or $0 \to 1 \,\ell^+ \ell^-$ meson decays, where we indicate the spin of the initial and final hadron,
$A_{\text{FB}}^\ell$ is sensitive to the product of Wilson coefficients $C_9 C_{10}$ rather than the sum of squares as the branching ratios. Therefore,  in terms of constraints, the latter provide ellipses, while $A_{\text{FB}}^\ell$ gives hyperbolas in the $C_9$-$C_{10}$ plane \cite{Ali:1994bf}. Further observables with complementary dependence on the Wilson coefficients arise from full angular analysis.

Already a rough  measurement of   $A_{\text{FB}}^\ell$  can signal new physics,
and rule out large classes of BSM models and non-SM flavor structures~\cite{Ali:1999mm}.
The Belle collaboration determined  the 
sign of  $A_{\text{FB}}^\ell$  in $b \to s$ transitions  based on $113.6 \pm 13$ events of $B \to K^* \ell^+ \ell^-$   decays \cite{Belle:2006gil}.
Since LHCb evidenced $\bar B_s \to \bar K^{* 0} \mu \mu$ with $38 \pm 12$ events~\cite{LHCb:2018rym},
one could expect that this crucial  test in the $b \to d$ sector  can be performed in the not too far future.

Let us further entertain a comparison with the $b\to s$ global fits. 
The latter can be related to $b \to d$ processes assuming that quark flavor violation is realized minimally -- induced by the known quark masses and mixing angles. 
Within the weak effective theory~\eqref{eq:hamiltonianNP}, 
this implies universal Wilson coefficients $C_i$, allowing them to be compared between the two fits. 
For the primed coefficients an additional factor of down to strange quark masses 
is involved, $C_i^{\prime \,(b \to d)}=(m_d/m_s)\,C_i^{\prime \, (b \to s)}$.
A disagreement between the resulting $b\to s$ and $b\to d$  Wilson coefficients would signal additional sources of quark flavor violation beyond the SM. 
In \cref{fig:2dfit_contours_C78,fig:2dfit_contours}, 
we show in addition to the $b \to d$ results (red)  the global fit results for the same scenarios using $b\to s$ input (blue),  see \cref{app:btosfit}. 
The fact that the SM (black star) is not included in the $b \to s$ preferred regions is a reflection of the present anomalies in this sector.

We observe as expected from the significantly more precise data in the $b \to s $ sector, 
that the latter is significantly stronger constrained than the $b \to d$ one. 
Due to the quark mass suppression, 
$m_d/m_s \sim 0.1$, 
the derived constraints on the primed coefficients  are even stronger, 
and result in narrow regions in the corresponding plots.
One also notices that the preferred region 
for $b\to s$  is inside the $b\to d$ one. 
Data are therefore consistent with the hypothesis of minimal quark flavor violation. 
To test this  further requires improved data on $b\to d$ transitions. The anticipated reach in $B_d  \to \mu^+ \mu^-$ from the HL-LHC \cite{DiCanto:2022icc} given in \cref{eq:C10future} probes the present MFV prediction in $C_{10}$, see
{\it e.g.},  the upper left plot of \cref{fig:2dfit_contours}.
 On the other hand,  there is a sizable window for flavorful new physics in $b \to d$ modes, and new physics can be just around the corner.
  Observing  for instance a wrong sign  $A_{\text{FB}}^\ell$  
 with the next round of data  would signal a breakdown of the SM, and of MFV.

\subsubsection{Four-dimensional fits}

We perform 4D fits (scenarios $H_{38}$ and $H_{39}$), where the numerical results are presented in \cref{tab:fitres_4d}. 
Results are consistent with those of the  1D and 2D scenarios, but exhibit larger uncertainties due to the larger number of parameters. 
This is to be expected as already good fits where obtained for 1D and 2D scenarios, 
so the additional degrees of freedom are not needed to improve the consistency with the data. 
For this reason, we do not show contour plots of the 4D fits. 
On the other hand, the more general situation with four independent couplings describes more general NP models.

\subsection{Interplay with $b \to d\, \nu \bar \nu$ and high-$p_T$\label{sec:smeft}}

For completeness, we compare our results with the limits extracted using other experimental information. 
To do so, it is convenient to introduce the following combinations of Wilson coefficients
\begin{align}
    \kappa_{L}^{bd\mu\mu}\,=\,C_9 -C_{10}~,~~ \kappa_{R}^{bd\mu\mu}\,=\,C_9^{\prime}-C_{10}^{\prime}~,
\end{align}
which project onto couplings to left-handed leptons. The methodology is based on \cite{Bause:2021cna}.  In \cref{tab:limitsonkappa}, we show the limits on $\kappa_{L(R)}^{bd\mu\mu}$ obtained in the 4D scenario $H_{38}$ $(C_9,C_9^\prime,C_{10},C_{10}^\prime)$, which update those from \cite{Bause:2021cna} by taking into account correlations. Also  given are  limits using data on rare $B$ decays into dineutrinos~\cite{Bause:2021cna}, as well as high-$p_T$ constraints~\cite{Fuentes-Martin:2020lea,Angelescu:2020uug}. The limits from the global fit obtained in this work are a factor 30 (20) stronger than the limits extracted from dineutrino modes (Drell-Yan processes).

\begin{table}[h!]
    \centering
 \resizebox{0.47\textwidth}{!}{
 \renewcommand{\arraystretch}{1.55}
    \begin{tabular}{c|c|c}
    \hline
    \hline
      \rowcolor{LightBlue}
     Data & $\kappa_{R}^{bd\mu\mu}$ & $\kappa_{L}^{bd\mu\mu}$ \\
     \hline
     \hline
        Rare $B$ decays to  dimuons & $[-7.2,7.2]$ & $[-15.0,-0.6]$ \\
    \hline
     Rare $B$ decays to dineutrinos &$210$ & $-$ \\
    \hline
     Drell-Yan  & $314$ & $314$ \\
    \hline
    \hline
    \end{tabular}
    }
    \caption{Limits on $\kappa_{R,L}^{bd\mu\mu}$ couplings. The second row gives the $1\sigma$ results from our global $b\to d\,\mu^+\mu^-$ fit  to $C_9,C_9^\prime,C_{10},C_{10}^\prime$  (scenario $H_{38}$). The third row displays the bound on $\kappa_{R}^{bd\mu\mu}$ from dineutrino modes using $SU(2)_L$ taken from Ref.~\cite{Bause:2021cna}. The last row displays upper limits extracted from high-$p_T$ data, that is, Drell-Yan processes~\cite{Fuentes-Martin:2020lea,Angelescu:2020uug}.}
    \label{tab:limitsonkappa}
\end{table}

\begin{figure}[h!]
    \centering
    \includegraphics[scale=0.5]{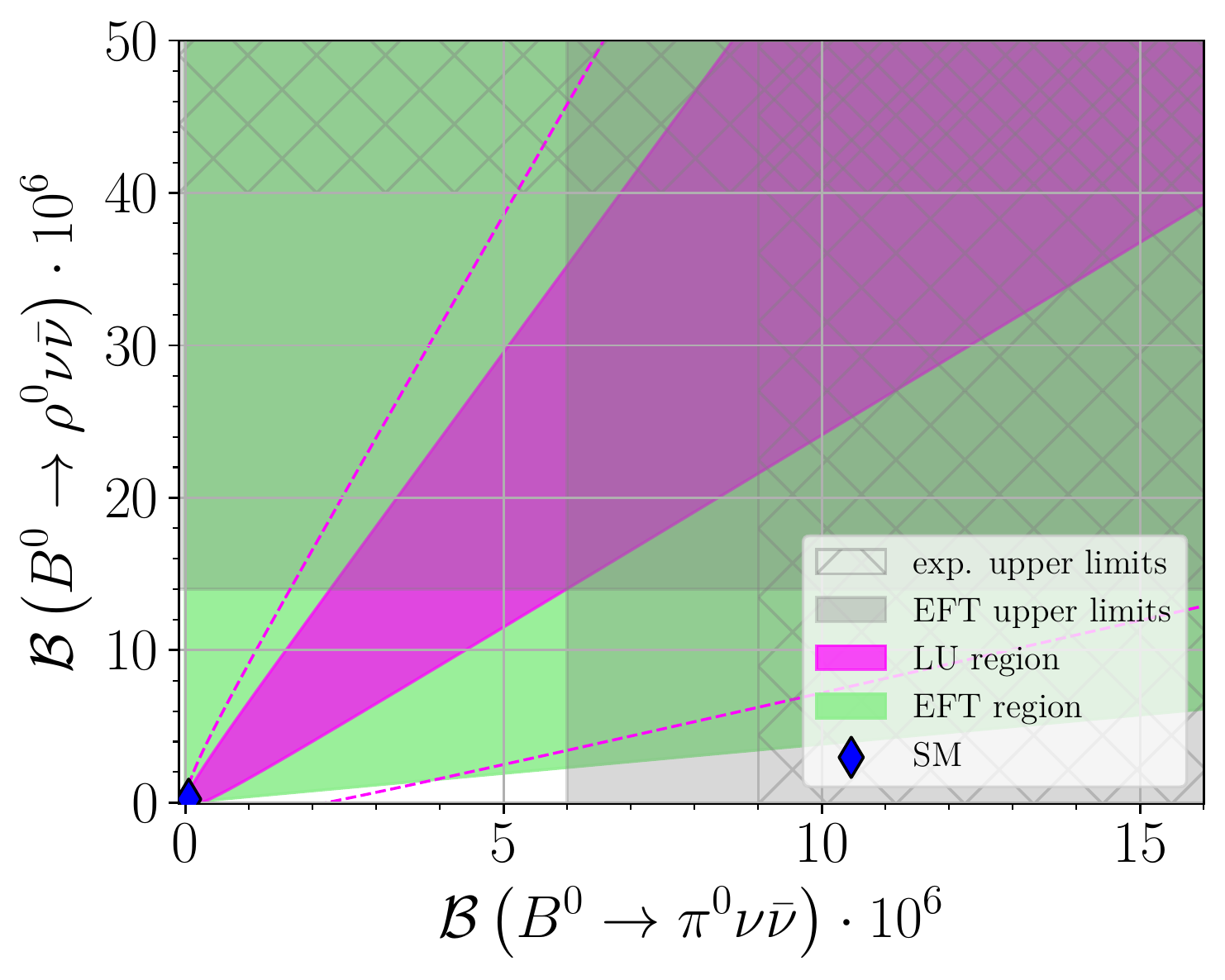}
    \caption{$\mathcal{B}(B^0\to\rho^0\nu\bar{\nu})$ versus $\mathcal{B}(B^0\to\pi^0\nu\bar{\nu})$. The region consistent with lepton universality is shown  at $1 \sigma$ (purple area)
    and $2 \sigma$ (purple dashed lines)
     using $SU(2)_L$ and $\kappa_R^{bd\mu\mu}=0.0\pm7.2$. SM predictions \eqref{eq:SMnunu} are shown as the blue diamond.
     The non-green area is outside the effective theory description. 
     Gray regions indicate experimental exclusion  limits. 
     See Ref.~\cite{Bause:2021cna} and text for details.}
        \label{fig:Brhovv_correlation_update}
\end{figure}

Applications of the FCNC $|\Delta b|=|\Delta d|=1$ global fits are flavor studies in  dineutrino modes \cite{Bause:2020auq}. Branching ratios of $b \to d\, \nu \bar \nu$ decays are related to lepton flavor-specific data using an $SU(2)_L$-link, that can be formulated in SMEFT. Constraints on dineutrinos derived from charged dileptons depend on whether lepton universality or charged lepton flavor violation holds, and can hence probe lepton flavor structure. This is demonstrated in \cref{fig:Brhovv_correlation_update}, 
where $\mathcal{B}(B^0\to\rho^0\nu\bar{\nu})$ versus $\mathcal{B}(B^0\to\pi^0\nu\bar{\nu})$ is displayed. 
We show the region consistent with lepton universality  at $1 \sigma$ (purple area) and $2 \sigma$ (purple dashed lines). 
The SM predictions (blue)~\cite{Bause:2021cna}
\begin{align} \label{eq:SMnunu}
\begin{split}
     \mathcal{B}(B^0 \to \pi^0 \nu \bar \nu)_{\text{SM}}&=(5.4\pm 0.6)\cdot 10^{-8},\\
    \mathcal{B}(B^0 \to \rho^0 \nu \bar \nu)_{\text{SM}}&=(22\pm 8)\cdot 10^{-8} , 
\end{split}
\end{align}
are well below the experimental limits (gray exclusion areas). 
The hatched ones are direct upper limits while the unhatched ones are derived from the presently strongest upper limits on the dineutrino modes, $ {\mathcal{B}}(B^+ \to \pi^+ \nu \bar \nu) < 1.4 \cdot 10^{-5}$ and $ {\mathcal{B}}(B^+ \to \rho^+ \nu \bar \nu)< 3 \cdot 10^{-5}$ (at 90\% CL)~\cite{Belle:2017oht}. 
If dineutrino branching ratios are observed in the future outside the purple cones lepton flavor universality is violated.

\Cref{fig:Brhovv_correlation_update} updates the corresponding plot from Ref.~\cite{Bause:2021cna} by taking into account the somewhat larger range of $\kappa_{R}^{bd\mu\mu}$ due to correlations in the global fit~\footnote{The ranges from \cite{Bause:2021cna} are
$\kappa_{L}^{bd\mu\mu}=-3 \pm 5$, $\kappa_{R}^{bd\mu\mu}=0 \pm 4$. For the correlation shown in \cref{fig:Brhovv_correlation_update} only $\kappa_{R}^{bd\mu\mu}$ matters.}. 
While the cones predicted assuming universality (purple) are consequently somewhat wider, this is a barely visible effect as the width is fully dominated by FF uncertainties, see Ref.~\cite{Bause:2021cna} for details.

\section{Conclusions \label{sec:con}}

We present the  first  model-independent analysis of  rare radiative and semileptonic $|\Delta b|=|\Delta d|=1$  processes based on the  branching ratios of $B^+ \to \pi^+ \mu^+\mu^-$, $B^0_s\to \bar{K}^{*0} \mu^+\mu^-$, $B^0\to\mu^+\mu^-$, and $B \to X_d \gamma$ decays.
The detailed numerical outcome of the global fits to Wilson coefficients of dipole $C_7^{(\prime)},C_8^{(\prime)}$  and four-fermion operators $C_9^{(\prime)},C_{10}^{(\prime)}$  is given in \cref{tab:fitres_1d,tab:fitres_2d,tab:fitres_4d}.
 
We find that data are consistent with the SM, but also leave sizable room for new physics. 
Contributions to Wilson coefficients can be order one in $b\to d$ processes, significantly larger than corresponding ranges in the $b \to s$ transitions, as shown in \cref{fig:2dfit_contours_C78,fig:2dfit_contours}. The preferred regions of the $b \to d$ Wilson coefficients (red areas), collapse to the smaller, included ones from $b\to s$ fits (blue areas) in models with minimal quark flavor violation, indicating consistency with MFV. 
To test this paradigm at  the level  of the current $b \to s$ prediction the $b \to d$  fit needs to be significantly improved.
This is within the HL-LHC reach of the $B_d \to \mu^+ \mu^-$  branching  for the Wilson coefficient $C_{10}$ \eqref{eq:C10future}.
In the meantime, since  flavorful new physics may be just around the corner, it  can show up in $b \to d$  angular distributions and could rule out MFV. 
 Note that the pattern of  branching ratios suppressed with respect to the SM in 
$b \to s\,\mu \mu$ decays
continues to hold for the $ b \to d\,\mu \mu$ ones, although within larger uncertainties, see  
\cref{fig:lhcbplot,eq:SMBtoKstar,eq:expBtoKstar}.

Due to several flat directions in parameter space but also sizable uncertainties, the likelihood for the semileptonic four-fermion operators is quite flat. 
Consequently, improving the fit is not just a matter of higher statistics, but also of adding observables sensitive to different combinations of Wilson coefficients than the branching ratios. 
Such observables  are  the very well-known leptonic forward-backward asymmetries in semileptonic decays of $b$-mesons or $b$-baryons, see \cref{eq:AFB}, $B_s^0 \to  \bar{K}^{*0} \,   \ell^+ \ell^-$, $B \to \rho  \,\ell^+ \ell^-$, $\Xi_b \to \Sigma\, \ell^+ \ell^-$ or $\Omega_b^- \to \Xi^-\,   \ell^+ \ell^-$ or a full angular analysis thereof including secondary hadronic decays. 
We point out here opportunities from $\Omega_b^- \to \Xi^- \,(\to \Lambda \pi^-)\,  \ell^+ \ell^-$, as the $\Xi^-\to \Lambda \,\pi^-$ decays self-analyzing. 
The decay $\Lambda_b \to n \,\ell^+ \ell^-$ is also sensitive to the $|\Delta b|=|\Delta d|=1$ couplings but challenging experimentally due to the neutron in the final state.
We stress that already an  early, first measurement of $A_{\rm FB}^\ell$, as in  \cite{Belle:2006gil},  provides key information to the global fit.
Observing, for instance, a  'wrong sign'-value would rule out the SM, and point to flavorful new physics beyond MFV~\cite{Ali:1999mm}.

Rare $b \to d \,\ell \ell$ induced decays can be studied at high luminosity flavor facilities, such as LHCb~\cite{Cerri:2018ypt}, Belle II~\cite{Belle-II:2018jsg}, and a possible future $e^+ e^- $-collider running at the $Z$~\cite{FCC:2018byv}. For distributions of the dilepton mass sufficient separation from resonances is advised, and results should be provided in such bins, ideally with correlations, and without prior extrapolation over veto regions.

Improved knowledge of hadronic form factors, including $B \to \rho$ tensor ones that, for instance, would allow obtaining useful constraints from $B\to \rho\, \gamma$ data, would be desirable. With more data, also fits to complex-valued Wilson coefficients probing CP violation could be performed.
  
Besides testing the SM and quark flavor patterns, results of  the $|\Delta b|=|\Delta d|=1$ fit are key input to a novel test of lepton flavor universality using the dineutrino modes $B\to \pi\, \nu \bar \nu$, $B\to \rho \,\nu \bar \nu$~\cite{Bause:2021cna}, 
shown in \cref{fig:Brhovv_correlation_update}. 
Further synergies from flavorful analyses in SMEFT can be anticipated.
  
\acknowledgments
 
We are happy to thank  Johannes Albrecht and Tom Blake for useful communication on the LHCb analysis of $B^0_s \to  \bar{K}^{*0} \, \mu^+ \mu^-$. This work is supported by the \textit{Studienstiftung des Deutschen Volkes} (MG) and the \textit{Bundesministerium f\"ur Bildung und Forschung} --BMBF (HG) under project number 05H21PECL2. This work was  performed in part (GH) at Aspen Center for Physics, which is supported by National Science Foundation grant PHY-1607611.
\bigskip

\textbf{Notes added:} A recent study \cite{Biswas:2022lhu}  performs a sensitivity study to NP in the $B \to \rho \,\ell \ell$ full angular distribution.

\begin{figure}
    \centering
    \includegraphics[scale=0.49]{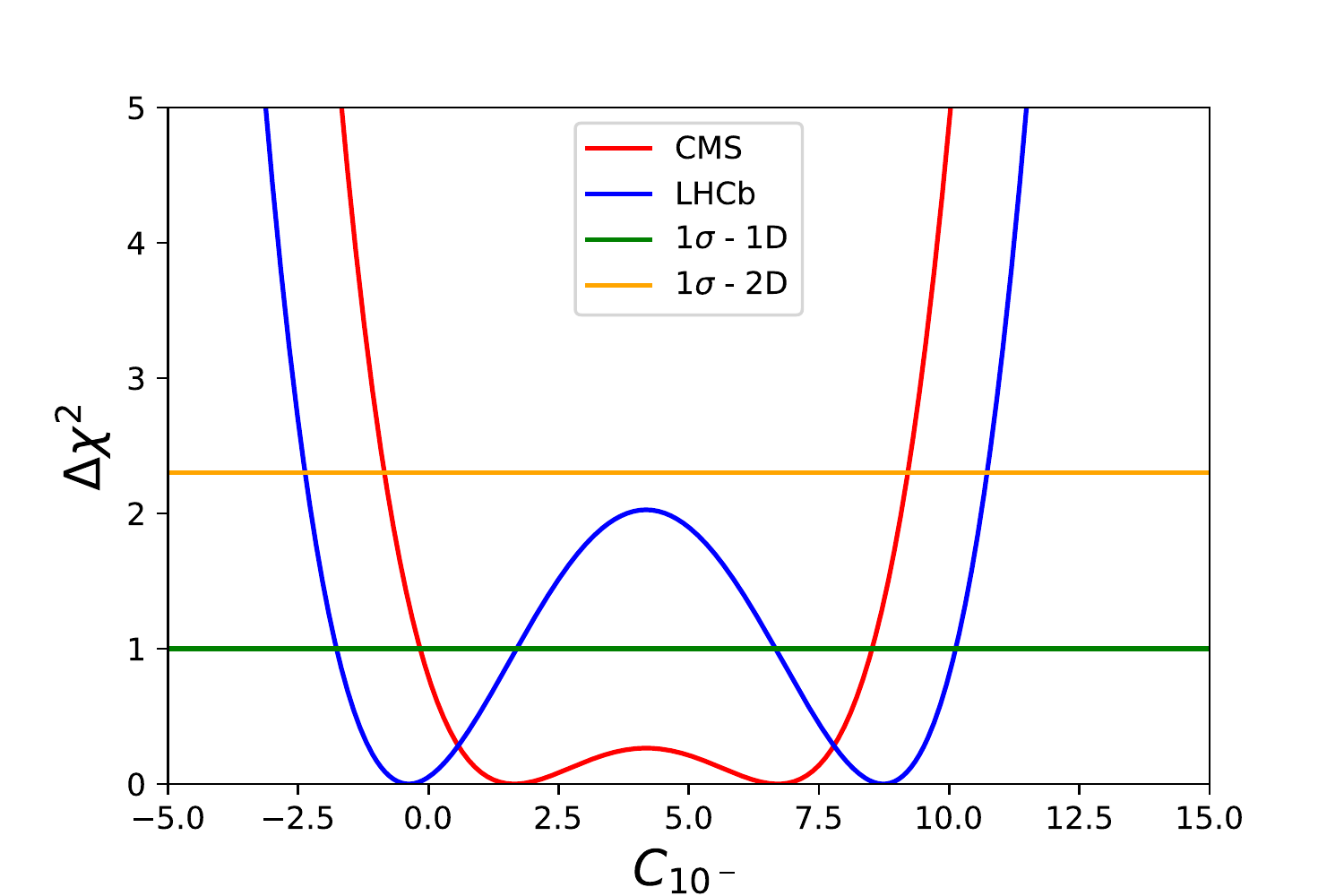}
    \caption{
    $\Delta\chi^2$ as function of $C_{10^-}$. 
    The plot shows the impact of LHCb~\cite{LHCb:2021awg} (blue) and CMS~\cite{CMS:2022mgd} (red) measurements on $C_{10^-}$ for the 1D (green) and 2D (orange) fits at $1\sigma$. }
    \label{fig:comparisonLHCbCMS}
\end{figure}

The CMS-collaboration reported recently $ \mathcal{B}(B^0\to\mu^+\mu^-) <1.9 \cdot 10^{-10}$ \cite{CMS:2022mgd}, which is consistent with LHCb's finding \eqref{eq:expBtomumu}
used in this work. We study the impact of the new CMS measurement in \cref{eq:ratioBtomumuexp,eq:C10mumu,eq:pscalar,eq:scalar} finding
\begin{align}
        &\frac{\mathcal{B}(B^0\to\mu^+\mu^-)_\text{exp}^{\text{CMS}}}{\mathcal{B}(B^0\to\mu^+\mu^-)_{\text{SM}}}\,=\,0.37\pm 0.71~,\\
        &-0.2 \lesssim C_{10^-} \lesssim 8.5~,\label{eq:C10mCMS}\\
        &-0.01 \lesssim C_{P^-} \lesssim 0.3~,\\
        &|C_{S^-}|  \lesssim 0.2~,
\end{align}
respectively. 
\Cref{fig:comparisonLHCbCMS} illustrates the impact of LHCb and CMS data on $C_{10^-}$ in both the 1D (green) and 2D (orange) fits, with a 1$\sigma$ uncertainty range.
Although the $B_{d,s} \to \mu^+ \mu^-$ data from both experiments have not yet been combined, such an analysis would be highly desirable for improving the $b\to d, b \to s$ fits, as evident from \cref{fig:comparisonLHCbCMS}. 
On one hand, this analysis would eliminate the current LHCb degeneracy of $C_{10^-}$ in the 1D fits, as seen in the intersection of the blue curve with the green horizontal line. 
On the other hand, combining the data from both experiments would greatly reduce the uncertainty of this parameter, as demonstrated by the asymptotic behavior of the $\Delta\chi^2$ functions, see \cref{fig:comparisonLHCbCMS}.

\appendix

\section{Covariance matrices}\label{app:covarianceObs}

Here, we describe the construction of the experimental and theoretical covariance matrices utilized in our analysis. We do not consider correlations between the experimental data and the theory inputs or vice-versa. Hence, the experimental and theoretical covariance matrices are added in \cref{eq:covmatrix}. In addition, we do not consider correlations between different observables except for the bins of differential branching ratios of $B^+\to\pi^+\mu^+\mu^-$  decays. These assumptions result in the following block-diagonal covariance matrix
\begin{align}
    V^{(X)}=\begin{pmatrix}
[V_{B\pi}^{(X)}]_{3\times 3} & 0_{3\times 1} & 0_{3\times 1} & 0_{3\times 1}\\
0_{1\times 3} & V_{B_s\bar{K}^*}^{(X)} & 0 & 0\\
0_{1\times 3} & 0 & V_{bd\mu}^{(X)} & 0\\
0_{1\times 3} & 0 & 0 & V_{bd\gamma}^{(X)}
\end{pmatrix},\,
\end{align}
where $V_{B\pi}^{(X)}$, $V_{B_s\bar{K}^*}^{(X)}$, $V_{bd\mu}^{(X)}$, and $V_{bd\gamma}^{(X)}$ with $X=\text{exp}\,(\text{th})$, are the experimental (theoretical) covariance matrices of the three binned branching fractions $\mathcal{B}^{(B\pi)}_{1,2,9}$, and the branching ratios $\mathcal{B}(B_s^0\,\to\,\bar{K}^{*0}\,\mu^+\mu^-)$, $\mathcal{B}(B^0\,\to\,\mu^+\mu^-)$, and $\mathcal{B}(\bar{B}\,\to\,X_d\,\gamma)$, respectively.

\subsection{$V^{(\text{exp})}$}

The LHCb collaboration~\cite{LHCb:2015hsa} does not provide the statistical and systematic correlation matrices for the block $[V_{B\pi}^{(\text{exp})}]_{3\times 3}$. Therefore,  we conservatively assume independent statistical uncertainties and maximally correlated systematic errors,
\begin{align}
[V_{B\pi}^{(\text{exp})}]_{ij}\,=\,\sigma_i^2\,\delta_{ij}\,+\,\epsilon_i\,\epsilon_j~,
\end{align}
where  the statistical $\sigma_i$ and systematic $\epsilon_i$ uncertainties are provided in \cref{tab:Bpi}. For the other entries $V^{(\text{exp})}_\alpha$ with $\alpha=B_s\bar{K}^*,\,bd\mu,\,bd\gamma$, 
we use $V^{(\text{exp})}_k=\sigma^2_\alpha$ 
where the experimental errors are taken from \cref{eq:expBtoKstar,eq:expBtomumu,eq:Exp_b_to_d_gamma}, respectively.

\subsection{$V^{(\text{th})}$}

We start with the block $[V_{B\pi}^{(\text{th})}]_{3\times 3}$,  where three sources of uncertainty (FFs, CKM matrix elements, and the variation of the short-distance scale $\mu_b$) are included as
\begin{align}
V_{B\pi}^{(\text{th})}\,=\,V_{B\pi}^{(\text{FFs})}\,+\,V_{B\pi}^{(\text{CKM})}\,+\,V_{B\pi}^{(\mu_b)}\,.  
\end{align}
The matrix $V_{B\pi}^{(\text{FFs})}$ can be written as
\begin{align}\label{eq:VBPi_FF}
\left[V_{B\pi}^{(\text{FFs})}\right]_{ij}\,=\,\left[\rho^{(\text{FFs})}_{B\pi}\right]_{ij}\,\sigma_{B\pi,\,i}^{(\text{FFs})}\,\sigma_{B\pi,\,j}^{(\text{FFs})}~,
\end{align}
where $\sigma_{B\pi,\,i}^{(\text{FFs})}$ is the FF uncertainty of $\mathcal{B}_i^{(B\pi)}$. In \cref{eq:VBPi_FF}, $[\rho^{(\text{FFs})}_{B\pi}]_{ij}$ denotes the correlation matrix for the FFs between $\mathcal{B}_i^{(B\pi)}$ and $\mathcal{B}_j^{(B\pi)}$. It can be expressed via the quantity $\Delta\mathcal{B}^{ (B\pi)}_{ij}=\mathcal{B}_i^{(B\pi)}+\mathcal{B}_j^{(B\pi)}$ with the uncertainty $\sigma_{\Delta B\pi,\,ij}^{(\text{FFs})}$, and we write
\begin{align}\label{eq:corrBpiFFs}
    [\rho^{(\text{FFs})}_{B\pi}]_{ij}\,=\,\frac{\left(\sigma_{\Delta B\pi,\,ij}^{(\text{FFs})}\right)^2\,-\,\left(\sigma_{B\pi,\,i}^{(\text{FFs})}\right)^2\,-\,\left(\sigma_{B\pi,\,j}^{(\text{FFs})}\right)^2}{2\,\left(\sigma_{B\pi,\,i}^{(\text{FFs})}\right)\,\left(\sigma_{B\pi,\,j}^{(\text{FFs})}\right)}~,
\end{align}
with
\begin{align}
(\sigma_{\Psi}^{(\text{FFs})})^2=\sum_{\mathcal{F},\mathcal{F}^\prime}\sum_{k,k^\prime}\frac{\partial \Psi}{\partial\alpha_k^{(\mathcal{F})}}\text{cov}\left(\alpha_k^{(\mathcal{F})},\alpha_{k^\prime}^{(\mathcal{F}^\prime)}\right)\frac{\partial \Psi}{\partial\alpha_{k^\prime}^{(\mathcal{F}^\prime)}}~,\nonumber
\end{align}
where $\Psi$ is a generic function that depends on the parameters $\alpha_{k}^{(\mathcal{F})}$ from the FF $\mathcal{F}$, see Ref.~\cite{Leljak:2021vte}. 

Since the main dependence on the CKM matrix elements factorizes, that is $\mathcal{B}_{i}^{(B\pi)}\,=\,\left|V_{tb}\,V_{td}^*\right|^2\times (...)$, we estimate this uncertainty as
\begin{align}
\sigma_{B\pi,\,i}^{(\text{CKM})}\,=\,\mathcal{B}_{i}^{(B\pi)}\,\frac{\delta(\left|V_{tb}\,V_{td}^*\right|^2)}{\left|V_{tb}\,V_{td}^*\right|^2}~,
\end{align}
where $\delta(\left|V_{tb}\,V_{td}^*\right|^2)$ refers to the uncertainty of $\left|V_{tb}\,V_{td}^*\right|^2\,=\,(7.3 \pm 0.4)\cdot 10^{-5}$~\cite{ParticleDataGroup:2022pth}. The covariance matrix $V_{B\pi}^{(\text{CKM})}$ is given by
\begin{align}\label{eq:VBPi_CKM}
   [V_{B\pi}^{(\text{CKM})}]_{ij}\,=\,  \sigma_{B\pi,\,i}^{(\text{CKM})}\, \sigma_{B\pi,\,j}^{(\text{CKM})}\,\delta_{ij}~.
\end{align}
For $V_{B\pi}^{(\mu_b)}$, we perform a variation of the short-distance scale $\mu_b$ between $3$ to $5\,\gev$ using Table~I from Ref.~\cite{Ali:2013zfa}, and estimate the uncertainty of $\mathcal{B}_i^{(B\pi)}$ as 
\begin{align}
    \sigma_{B\pi,\,i}^{(\mu_b)}\,=\,\frac{\left|\mathcal{B}_{i}^{(B\pi)}(5\,\gev)-\mathcal{B}_{i}^{(B\pi)}(3\,\gev)\right|}{2}~.
\end{align}
We  obtain
\begin{align}\label{eq:VBPi_mub}
    [V_{B\pi}^{(\mu_b)}]_{ij}\,=\,\sigma_{B\pi,\,i}^{(\mu_b)}\,\sigma_{B\pi,\,j}^{(\mu_b)}\,\delta_{ij}~.
\end{align}
Adding \cref{eq:VBPi_FF,eq:VBPi_CKM,eq:VBPi_mub}, we obtain  $[V_{B\pi}^{(\text{th})}]_{ij}$. For the correlation matrices $V_{B_s\bar{K}^*}^{(\text{th})}$, $V_{bd\mu}^{(\text{th})}$, and $V_{bd\gamma}^{(\text{th})}$, we follow a similar procedure as the one for $V_{B\pi}^{(\text{th})}$.

\section{Cuts in $B_s^0\to \bar K^{*0}\,\mu^+\mu^-$}\label{app:resonances}

Here we estimate the uncertainty due to contributions from  $J/\psi$ and $\psi^\prime$ resonances in the $B_s^0\to \bar K^{*0}\,\mu^+\mu^-$ branching ratio, which corresponds to the last uncertainty in \cref{eq:SMBtoKstar}. This is necessary to interpret the data~\cite{LHCb:2018rym}, which is provided only for a single large bin encompassing the resonance regions: while kinematic cuts have been applied to remove the bulk of the $J/\psi$ and $\psi^\prime$,
$q^2\in[8,\,11]\,\gev^2$ ($J/\psi$) and $q^2\in[12.5,\,15]\,\gev^2$ ($\psi^\prime$), their tails enter the signal regions and affect the interpolation between them. We stress that this procedure introduces some degree of model dependence and additional uncertainty and 
can be avoided once data in theory-friendly bins are available. To make progress  we model these effects using a constant-width Breit-Wigner approximation, 
\begin{align}\label{eq:BreitWigner}
    C_9^{\text{res}}(q^2)=\sum_{R}\frac{a_R\,\text{e}^{\text{i}\,\delta_R}}{q^2-m_R^2+\text{i}\,m_R\,\Gamma_R}~,
\end{align}
with `fudge factor' $a_R>0$ and $R=J/\psi,\,\psi^\prime$, and $m_R,\Gamma_R$ the resonance mass and width. This simple ansatz
serves to estimate the uncertainty of the interpolation but we note that a more sophisticated prescription~\cite{Kruger:1996cv} could be used for the resonance shapes. Close to $q^2\approx m_R^2$, the differential branching ratio is dominated by the resonance $R$, that is 
\begin{align}\label{eq:diff}
    \frac{\text{d}\,\mathcal{B}(B_s^0\to \bar K^{*0}\,\mu^+\mu^-)}{\text{d}\,q^2}\,\propto \frac{f(q^2)\,a_R^2}{(q^2-m_R^2)^2\,+\,m_R^2\,\Gamma_R^2}~.
\end{align}
Here, $f(q^2)$ is a function obtained from \cref{eq:diffBtoKstar,eq:functionBtoKstar,eq:BtoKstarABC}. In the limit $\Gamma_R/m_R\to 0$, \cref{eq:diff} becomes a $\delta$-distribution, and integration over $q^2$ yields
\begin{align}\label{eq:Brres}
    \mathcal{B}(B_s^0\to \bar K^{*0}\,\mu^+\mu^-)\, \propto \,\frac{\pi\,a_R^2\,f(m_R^2)}{m_R\,\Gamma_R}~.
\end{align}
In this limit, we can only produce the resonance $R$ on-shell, which later on decays to muons
\begin{align}\label{eq:Brres1}
     \mathcal{B}(B_s^0\to \bar K^{*0}\,\mu^+\mu^-)=\mathcal{B}(B_s^0\to \bar K^{*0}\,R)\,\mathcal{B}(R\to\mu^+\mu^-)~.
\end{align}
Using \cref{eq:Brres,eq:Brres1}, together with the experimental information on $m_R$, $\Gamma_R$, $\mathcal{B}(B_s^0\to \bar K^{*0}\,R)$, and $\mathcal{B}(R\to\mu^+\mu^-)$~\cite{ParticleDataGroup:2022pth},
one  extracts   $a_R$ as
\begin{align}
    a_R\,=\,\sqrt{\frac{m_R\,\Gamma_R\,\mathcal{B}(B_s^0\to \bar K^{*0}\,R)\,\mathcal{B}(R\to\mu^+\mu^-)}{\pi\,f(m_R^2)}}~,
\end{align}
resulting in the fudge factors
\begin{align}
\begin{split}
    a_{J/\psi}\,&=\,1.89\pm0.09~,\\
    a_{\psi^\prime}\,&=\,1.20\pm0.10~,
\end{split}
\end{align}
close to their naive factorization limit $a_R \simeq 1$. The only remaining unknowns  in \cref{eq:BreitWigner} are the phases $\delta_{R}\in[0,\,2\pi]$, and represent the main source of uncertainty.

\begin{figure}[b!]
    \centering
    \includegraphics[width=\columnwidth]{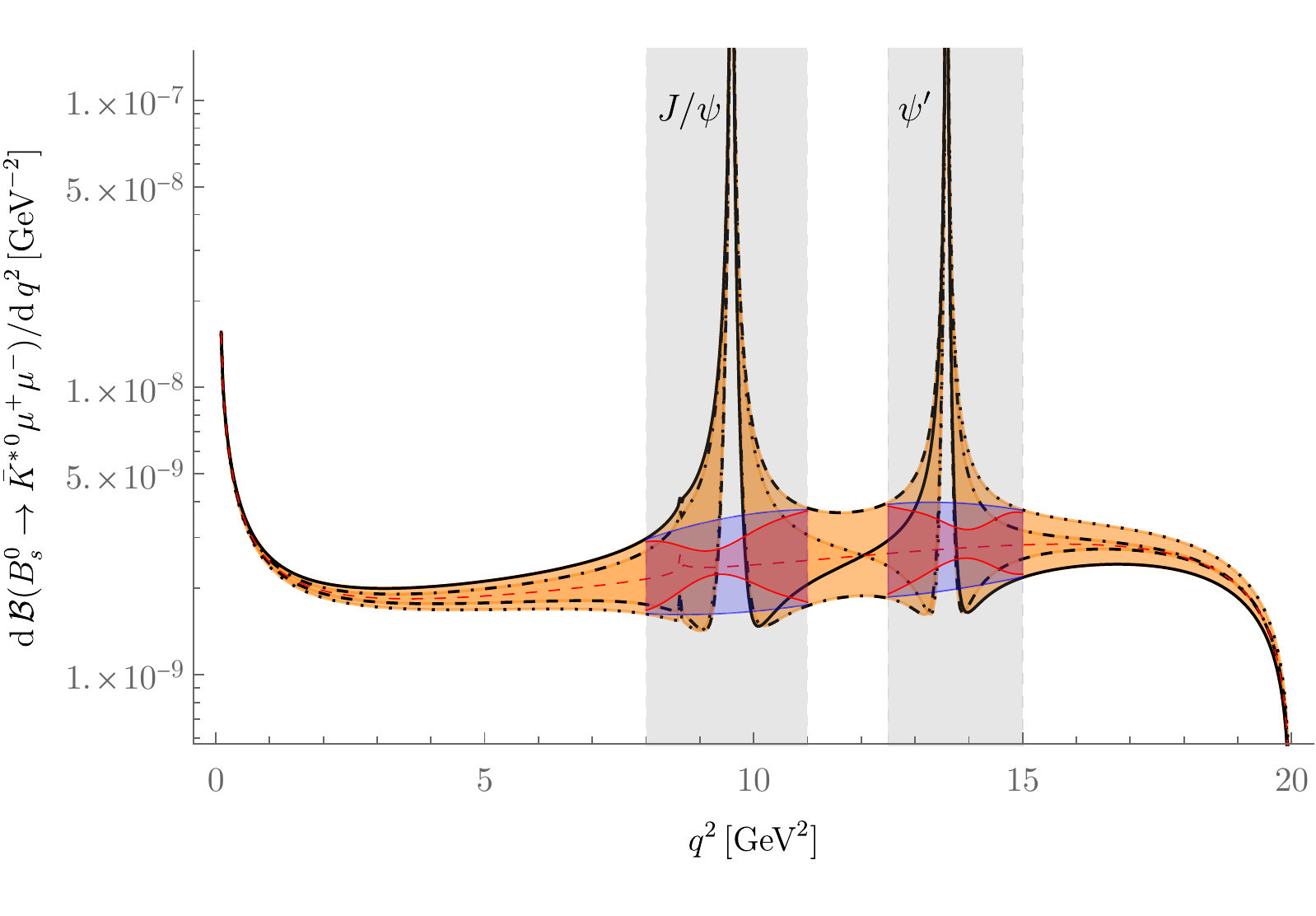}
    \caption{Differential branching ratio of $B^0_s\to \bar{K}^{*0} \mu^+\mu^-$ as a function of $q^2$. The vertical gray bands around the $J/\Psi$ and the $\psi^\prime$ correspond to the regions vetoed by LHCb from the branching ratio measurement~\cite{LHCb:2018rym}. The red dashed line depicts the central value of the non-resonant SM contribution. The orange region illustrates the SM behavior with both the non-resonant and the resonant contributions, obtained by scanning independently over both relative phases $\delta_{J/\psi,\psi^\prime}\in[0,\,2\pi]$.  Details on the blue and red areas in the veto regions are given in the text.} \label{fig:resonance}
\end{figure}

\Cref{fig:resonance} displays the differential branching ratio $\text{d}\mathcal{B}(B_s^0\to \bar K^{*0}\,\mu^+\mu^-)/\text{d}q^2$. The red dashed line depicts the central value of the non-resonant contribution in the SM. The orange region represents the SM prediction with both the non-resonant and the resonant contributions, obtained by scanning over both phase shifts $\delta_{J/\psi,\psi^\prime}\in[0,\,2\pi]$. 
The LHCb~\cite{LHCb:2018rym} veto regions around the $J/\psi$ and $\psi^\prime$ resonances are shown as gray-shaded bands. In addition, \cref{fig:resonance} shows the central values for those phase shifts that give the maximum/minimum differential branching ratio using  the LHCb $q^2$-cuts:
\begin{itemize}
    \item $q^2=8\,\gev^2$: maximum ($\delta_{J/\psi}=\pi,\,\delta_{\psi^\prime}=\pi$, black solid line) and minimum ($\delta_{J/\psi}=0,\,\delta_{\psi^\prime}=0$, black dotted line),
    \item $q^2=11,\,12.5\,\gev^2$: maximum ($\delta_{J/\psi}=0,\,\delta_{\psi^\prime}=\pi$, black dashed line) and minimum ($\delta_{J/\psi}=\pi,\,\delta_{\psi^\prime}=0$, black dot-dashed line),
    \item $q^2=15\,\gev^2$: maximum ($\delta_{J/\psi}=0,\,\delta_{\psi^\prime}=0$, black dotted line) and minimum ($\delta_{J/\psi}=\pi,\,\delta_{\psi^\prime}=\pi$, black solid line).
\end{itemize}
To make use of the data, we exclude the veto regions and interpolate between them and give a conservative estimation of this uncertainty. 
In \cref{fig:resonance}, we learn that the effect from resonances tends to cancel between the signal bins [0.1,8], [11,12.5], [15,19] $\gev^2$  for fixed values of $\delta_R$. Therefore the resonance contribution on the full bin [0.1,19] $\gev^2$ is dominated by the regions vetoed by the experiment. 
We  numerically cross-checked this assumption, which yields up to $\order{10^{-10}}$ corrections, compared to $\mathcal{B}_{\text{total}}\sim \order{10^{-8}}$, which is two orders of magnitude larger.
We hence obtain the residual (interpolation) uncertainty of the resonances $\varepsilon_{\text{model}}$  from  the resonance (veto) regions, as
\begin{align*}    \mathcal{B}_{[8,\,11]}^{\text{R}+\text{NR}}+\mathcal{B}_{[12.5,\,15]}^{\text{R}+\text{NR}}\longrightarrow \mathcal{B}_{[8,\,11]}^{\text{NR}}+\mathcal{B}_{[12.5,\,15]}^{\text{NR}}\,+\,\varepsilon_{\text{model}}~.
\end{align*}
It is computed in the following way: For each resonance region ([8,11]$\,\text{GeV}^2$ and [12.5,15]$\,\text{GeV}^2$) we consider three points to the left and three points to the right of the intervals corresponding to the minimal/maximal value of $\delta_R$, and then perform an interpolation between them (blue and red solid lines). Allowing for different strong phases on both sides of the veto intervals (blue region in \cref{fig:resonance}) gives a more conservative uncertainty estimation than assuming the same strong phases (red region in \cref{fig:resonance}), that is, the red area is inside the blue one. In the fits we employ the more conservative approach and assume that the strong phases are not correlated and take maximum/minimum values on both sides, corresponding to the blue area. The resulting symmetrized uncertainty of $\varepsilon_{\text{model}}$  is obtained as
\begin{align}\label{eq:resonanceuncer}
    \varepsilon_{\text{model}}\,=\,\sqrt{(\delta\varepsilon_{J/\psi})^2+(\delta\varepsilon_{\psi^\prime})^2}\,=\,3.6\cdot10^{-9}~,
\end{align}
corresponding to the last uncertainty in \cref{eq:SMBtoKstar}, with
\begin{align*}
    \delta\varepsilon_{J/\psi}\,&=\,\frac{1}{2}\int_{8 \text{GeV}^2}^{11 \text{GeV}^2}\text{d}q^2\,\left(\frac{\text{d}\,\mathcal{B}_{\text{max}}^{[8,11]}}{\text{d}\,q^2}-\frac{\text{d}\,\mathcal{B}_{\text{min}}^{[8,11]}}{\text{d}\,q^2}\right),\\
    \delta\varepsilon_{\psi^\prime}\,&=\,\frac{1}{2}\int_{12.5 \text{GeV}^2}^{15 \text{GeV}^2}\text{d}q^2\,\left(\frac{\text{d}\,\mathcal{B}_{\text{max}}^{[12.5,15]}}{\text{d}\,q^2}-\frac{\text{d}\,\mathcal{B}_{\text{min}}^{[12.5,15]}}{\text{d}\,q^2}\right)~.
\end{align*}
For comparison, the model uncertainty on the interpolation for the red region is smaller, $\varepsilon_{\text{model}} =2.2\cdot 10^{-9}$. We also note that the fraction of the full (after cuts and interpolation)  to the signal branching ratio is $1.33 \pm 0.24$, consistent  but with larger uncertainty that the one employed by LHCb from the $B^0 \to K^{0*} \mu \mu$ analysis, $1.532 \pm 0.001\,(\text{stat}) \pm 0.010\,(\text{syst})$ \cite{LHCb:2016ykl}.

\section{Global $b\to s$ fits}\label{app:btosfit}

In this appendix, we provide the results of the global fit to $b\to s \mu^+\mu^-,\gamma$ data including the
update on $ \mathcal{B}(B_s^0\to\mu^+\mu^-)$
from the CMS-collaboration~\cite{CMS:2022mgd}.
We follow the same approach as Ref.~\cite{Bause:2021cna}, where the python package
\flavio~\cite{Straub:2018kue} was employed, and refer there for details.  
We do not include the $R_K$ and $R_{K^*}$ measurements from LHCb~\cite{LHCb:2022zom,LHCb:2022qnv} since
these observables also include electron modes, and would require lepton-specific fits.
We recall that dielectron modes are presently much less probed experimentally than dimuon ones and have therefore also not been included in the $b \to d $ fits presented in this work.

\begingroup
\renewcommand*{\arraystretch}{0.9}
\begin{table}[h!]
 \centering
 \begin{tabular}{c|c|c|c|c}
  \hline
  \hline
  \rowcolor{LightBlue}
  scenario  & fit parameter & best fit ($1\sigma$)  &$\chi^2/\text{dof}$ & Pull$_{H_i}$ \\ 
  \hline
  \hline
   $H_{3}$ & $C_9$ &  $-0.85\pm0.21$ & 1.03 & 3.57  \\
   $H_{10}$ & $C_9=-C_{10}$ &  $-0.50\pm0.13$ & 1.01 & 4.07  \\
   $H_{17}$ & $(C_7,C_8)$ &  $(0.01\pm0.03,-0.25\pm0.45)$ & 1.11 & 0.30  \\
   $H_{20}$ & $(C_7,C_7^\prime)$ &  $(-0.007\pm0.014,0.02\pm0.02)$ & 1.11 & 0.62  \\
   $H_{23}$ & $(C_9,C_{10})$ &  $(-1.03\pm0.21,0.41\pm0.13)$ & 0.97 & 4.44  \\
   $H_{25}$ & $(C_9,C_9^\prime)$ &  $(-0.86\pm0.20,0.54\pm0.27)$ & 1.01 & 3.69  \\
   $H_{26}$ & $(C_9,C_{10}^\prime)$ &  $(-1.09\pm0.19,-0.37\pm0.10)$ & 0.95 & 4.85  \\
   $H_{28}$ & $(C_9^\prime,C_{10})$ &  $(0.38\pm0.23,0.30\pm0.13)$ & 1.06 & 2.60  \\
   $H_{29}$ & $(C_{10},C_{10}^\prime)$ &  $(0.22\pm0.18,-0.10\pm0.15)$ & 1.07 & 2.15  \\
   $H_{32}$ & $(C_9^\prime,C_{10}^\prime)$ &  $(0.24\pm0.25,-0.18\pm0.11)$ & 1.08 & 2.04  \\
  \hline  
  \hline
  \end{tabular}
  \caption{Fit results of Wilson coefficients $C_i^{(b\to s)}$ to $b\to s \mu^+\mu^-,\gamma$ data for some exemplary 1D and 2D scenarios at the scale $\mu_b$. Best-fit values with its $1\sigma$ uncertainties are displayed in the third column. 
  We also provide the $\chi^2/\text{dof}$ value and respective pull from the SM hypothesis.
  }
  \label{tab:btosfit}
\end{table}
\endgroup

\clearpage


\bibliographystyle{jhep}
\bibliography{ref_btod.bib}

\end{document}